\documentstyle[epsf,12pt]{article}

  \def\pp{{\mathchoice
          {%displaystyle
              \kern 1pt%
              \raise 1pt
              \vbox{\hrule width5pt height0.4pt depth0pt
                    \kern -2pt
                    \hbox{\kern 2.3pt
                          \vrule width0.4pt height6pt depth0pt
                          }
                    \kern -2pt
                    \hrule width5pt height0.4pt depth0pt}%
                    \kern 1pt
           }
            {%textstyle
              \kern 1pt%
              \raise 1pt
              \vbox{\hrule width4.3pt height0.4pt depth0pt
                    \kern -1.8pt
                    \hbox{\kern 1.95pt
                          \vrule width0.4pt height5.4pt depth0pt
                          }
                    \kern -1.8pt
                    \hrule width4.3pt height0.4pt depth0pt}%
                    \kern 1pt
            }
            {%scriptstyle
              \kern 0.5pt%
              \raise 1pt
              \vbox{\hrule width4.0pt height0.3pt depth0pt
                    \kern -1.9pt  %[e]=0.15pt
                    \hbox{\kern 1.85pt
                          \vrule width0.3pt height5.7pt depth0pt
                          }
                    \kern -1.9pt
                    \hrule width4.0pt height0.3pt depth0pt}%
                    \kern 0.5pt
            }
            {%scriptscriptstyle
              \kern 0.5pt%
              \raise 1pt
              \vbox{\hrule width3.6pt height0.3pt depth0pt
                    \kern -1.5pt
                    \hbox{\kern 1.65pt
                          \vrule width0.3pt height4.5pt depth0pt
                          }
                    \kern -1.5pt
                    \hrule width3.6pt height0.3pt depth0pt}%
                    \kern 0.5pt%}
            }
        }}

  \def\mm{{\mathchoice
                       {%displaystyle
                             \kern 1pt
               \raise 1pt    \vbox{\hrule width5pt height0.4pt depth0pt
                                  \kern 2pt
                                  \hrule width5pt height0.4pt depth0pt}
                             \kern 1pt}
                       {%textstyle
                            \kern 1pt
               \raise 1pt \vbox{\hrule width4.3pt height0.4pt depth0pt
                                  \kern 1.8pt
                                  \hrule width4.3pt height0.4pt depth0pt}
                             \kern 1pt}
                       {%scriptstyle
                            \kern 0.5pt
               \raise 1pt
                            \vbox{\hrule width4.0pt height0.3pt depth0pt
                                  \kern 1.9pt
                                  \hrule width4.0pt height0.3pt depth0pt}
                            \kern 1pt}
                       {%scriptscriptstyle
                           \kern 0.5pt
             \raise 1pt  \vbox{\hrule width3.6pt height0.3pt depth0pt
                                  \kern 1.5pt
                                  \hrule width3.6pt height0.3pt depth0pt}
                           \kern 0.5pt}
                       }}

\catcode`@=11
\def\un#1{\relax\ifmmode\@@underline#1\else
        $\@@underline{\hbox{#1}}$\relax\fi}
\catcode`@=12

\let\du=\du                     % dot-under
\def\a{\alpha}
\def\b{\beta}
\def\c{\chi}
\def\d{\delta}

\def\f{\phi}
\def\g{\gamma}

\def\j{\psi}

\def\l{\lambda}
\def\m{\mu}
\def\n{\nu}
\def\o{\omega}
\def\p{\pi}
\def\q{\theta}

\def\s{\sigma}
\def\t{\tau}

\def\x{\xi}
\def\z{\zeta}
\def\D{\Delta}

\def\G{\Gamma}

\def\L{\Lambda}

\def\S{\Sigma}

\def\ve{\varepsilon}

\def\cf{{\cal F}}

\def\ch{{\cal H}}

\def\ck{{\cal K}}

\def\cm{{\cal M}}

\def\bo{{\raise-.5ex\hbox{\large$\Box$}}}               % D'Alembertian
\def\pa{\partial}                                       % curly d
\def\de{\nabla}                                         % del
\def\pr{\prod}                                          % product
\def\TH{{\raise.2ex\hbox{$\displaystyle \bigodot$}\mskip-4.7mu \llap H \;}}
\def\face{{\raise.2ex\hbox{$\displaystyle \bigodot$}\mskip-2.2mu \llap {$\ddot
        \smile$}}}                                      % happy face
\def\dg{\sp\dagger}                                     % hermitian conjugate
\def\sp#1{{}^{#1}}                              % superscript (unaligned)
\def\Bar#1{\overline{#1}}                       % big bar
\def\VEV#1{\left\langle #1\right\rangle}        % < >
\def\abs#1{\left| #1\right|}                    % | |
\def\leftrightarrowfill{$\mathsurround=0pt \mathord\leftarrow \mkern-6mu
        \cleaders\hbox{$\mkern-2mu \mathord- \mkern-2mu$}\hfill
        \mkern-6mu \mathord\rightarrow$}
\def\dvec#1{\vbox{\ialign{##\crcr
        \leftrightarrowfill\crcr\noalign{\kern-1pt\nointerlineskip}
        $\hfil\displaystyle{#1}\hfil$\crcr}}}           % <--> accent
\def\dt#1{{\buildrel {\hbox{\LARGE .}} \over {#1}}}     % dot-over for sp/sb
\def\frac#1#2{{\textstyle{#1\over\vphantom2\smash{\raise.20ex
        \hbox{$\scriptstyle{#2}$}}}}}                   % fraction
\def\sfrac#1#2{{\vphantom1\smash{\lower.5ex\hbox{\small$#1$}}\over
        \vphantom1\smash{\raise.4ex\hbox{\small$#2$}}}} % alternate fraction
\def\bfrac#1#2{{\vphantom1\smash{\lower.5ex\hbox{$#1$}}\over
        \vphantom1\smash{\raise.3ex\hbox{$#2$}}}}       % "
\def\afrac#1#2{{\vphantom1\smash{\lower.5ex\hbox{$#1$}}\over#2}}    % "
\def\[{\lfloor{\hskip 0.35pt}\!\!\!\lceil}
\def\]{\rfloor{\hskip 0.35pt}\!\!\!\rceil}

\def\du#1#2{_{#1}{}^{#2}}
\def\ud#1#2{^{#1}{}_{#2}}

\def\ha{{\fracmm12}}
\def\tr{{\rm tr}}

\def\un{\underline}
\def\fracmm#1#2{{{#1}\over{#2}}}

\def\low#1{{\raise -3pt\hbox{${\hskip 0.75pt}\!_{#1}$}}}

\def\Dot#1{\buildrel{_{_{\hskip 0.01in}\bullet}}\over{#1}}
\def\dt#1{\Dot{#1}}

\def\sbar#1{\stackrel{*}{\Bar{#1}}}

\newskip\humongous \humongous=0pt plus 1000pt minus 1000pt
\def\caja{\mathsurround=0pt}
\def\eqalign#1{\,\vcenter{\openup2\jot \caja
        \ialign{\strut \hfil$\displaystyle{##}$&$
        \displaystyle{{}##}$\hfil\crcr#1\crcr}}\,}
\newif\ifdtup

\def\pl#1#2#3{Phys.~Lett.~{\bf {#1}B} (19{#2}) #3}
\def\np#1#2#3{Nucl.~Phys.~{\bf B{#1}} (19{#2}) #3}
\def\prl#1#2#3{Phys.~Rev.~Lett.~{\bf #1} (19{#2}) #3}
\def\pr#1#2#3{Phys.~Rev.~{\bf D{#1}} (19{#2}) #3}
\def\cqg#1#2#3{Class.~and Quantum Grav.~{\bf {#1}} (19{#2}) #3}

\def\ibid#1#2#3{{\it ibid.}~{\bf {#1}} (19{#2}) #3}

\parskip=\medskipamount                 % space between paragraphs (LaTeX)
\thicklines                         % thick straight lines for pictures (LaTeX)

\begin{document}

% =========================== UH title page ==========================

\thispagestyle{empty}               % no heading or foot on title page (LaTeX)

\def\border{                                            % UH border
        \setlength{\unitlength}{1mm}
        \newcount\xco
        \newcount\yco
        \xco=-24
        \yco=12
        \begin{picture}(140,0)
        \put(-20,11){\tiny Institut f\"ur Theoretische Physik Universit\"at
Hannover~~ Institut f\"ur Theoretische Physik Universit\"at Hannover~~
Institut f\"ur Theoretische Physik Hannover}
        \put(-20,-241.5){\tiny Institut f\"ur Theoretische Physik Universit\"at
Hannover~~ Institut f\"ur Theoretische Physik Universit\"at Hannover~~
Institut f\"ur Theoretische Physik Hannover}
        \end{picture}
        \par\vskip-8mm}

\def\headpic{                                           % UH heading
        \indent
        \setlength{\unitlength}{.8mm}
        \thinlines
        \par
        \begin{picture}(29,16)
        \put(75,16){\line(1,0){4}}
        \put(80,16){\line(1,0){4}}
        \put(85,16){\line(1,0){4}}
        \put(92,16){\line(1,0){4}}

        \put(85,0){\line(1,0){4}}
        \put(89,8){\line(1,0){3}}
        \put(92,0){\line(1,0){4}}

        \put(85,0){\line(0,1){16}}
        \put(96,0){\line(0,1){16}}
        \put(92,16){\line(1,0){4}}

        \put(85,0){\line(1,0){4}}
        \put(89,8){\line(1,0){3}}
        \put(92,0){\line(1,0){4}}

        \put(85,0){\line(0,1){16}}
        \put(96,0){\line(0,1){16}}
        \put(79,0){\line(0,1){16}}
        \put(80,0){\line(0,1){16}}
        \put(89,0){\line(0,1){16}}
        \put(92,0){\line(0,1){16}}
        \put(79,16){\oval(8,32)[bl]}
        \put(80,16){\oval(8,32)[br]}

        \end{picture}
        \par\vskip-6.5mm
        \thicklines}

\border\headpic {\hbox to\hsize{
\vbox{\noindent DESY 97 -- 199  \hfill hep-th/9710085: October 1997 \\
ITP--UH--26/97 \hfill                  in final form: November 1997 }}}

\noindent
\vskip1.3cm
\begin{center}

{\Large\bf       ON EXACT SOLUTIONS TO QUANTUM 
\vglue.1in          $N=2$ GAUGE THEORIES~\footnote{Supported in part by the 
`Deutsche Forschungsgemeinschaft' and the NATO grant CQG 930789}}\\
\vglue.3in

Sergei V. Ketov \footnote{
On leave of absence from:
High Current Electronics Institute of the Russian Academy of Sciences,
\newline ${~~~~~}$ Siberian Branch, Akademichesky~4, Tomsk 634055, Russia}

{\it Institut f\"ur Theoretische Physik, Universit\"at Hannover}\\
{\it Appelstra\ss{}e 2, 30167 Hannover, Germany}\\
{\sl ketov@itp.uni-hannover.de}
\end{center}
\vglue.2in
\begin{center}
{\Large\bf Abstract}
\end{center}

Exact solutions to the low-energy effective action (LEEA) of the
four-dimensional ($4d$), $N=2$ supersymmetric gauge theories with matter
(including $N=2$ super-QCD) are discussed from the three different viewpoints:
(i) instanton calculus, (ii) $N=2$ harmonic superspace, and (iii) M theory. 
The emphasis is made on the foundations of all three approaches and their 
relationship.

\newpage

\begin{center}
{\bf CONTENT}
\end{center}

\noindent
1. Introduction . . . . . . . . . . . . . . . . . . . . . . . . . . . . . . .
. . . . . . . . . \hfill{3}

1.1. Motivation . . . . . . . . . . . . . . . . . . . . . . . . . . . . . . . 
. . . . . . . \hfill{4}

1.2. Setup  . . . . . . . . . . . . . . . . . . . . . . . . . . . . . . . . . 
. . . . . . . . \hfill{6}

\noindent
2. Gauge LEEA in the Coulomb branch . . . . . . . . . . . . . . . . . . . . . 
. . . . \hfill{10}

2.1. On the instanton calculations . . . . . . . . . . . . . . . . . . . . . 
. . . . . . \hfill{11}

2.2. Seiberg-Witten curve . . . . . . . . . . . . . . . . . . . . . . . . . . 
. . . . . . \hfill{13}

\noindent
3. Hypermultiplet LEEA in the Coulomb branch  . . . . . . . . . . . . . . . . 
. . . . \hfill{14}

3.1. $N=2$ harmonic superspace  . . . . . . . . . . . . . . . . . . . . . . 
. . . . . . \hfill{15}

3.2. Induced Taub-NUT metric or KK-monopole . . . . . . . . . . . . . . . . . .
\hfill{19}

3.3. Duality transformation and $N=2$ tensor multiplet  . . . . . . . . . . . 
. .\hfill{23}

\noindent 
4. Brane technology . . . . . . . . . . . . . . . . . . . . . . . . . . . . . 
. . . . . . . . \hfill{27}

4.1. M-theory resolution  . . . . . . . . . . . . . . . . . . . . . . . . . .
. . . . . . . \hfill{29}

4.2. Relation to the HSS results and S-duality  . . . . . . . . . . . . . . . 
. . . . \hfill{31}

\noindent
5. On the next-to-leading-order correction to the gauge LEEA . . . . . . . . . 
. . . \hfill{32}

\noindent
6. Hypermultiplet LEEA in the Higgs branch  . . . . . . . . . . . . . . . . . 
. . . . . \hfill{35}

\noindent
7. Conclusion . . . . . . . . . . . . . . . . . . . . . . . . . . . . . . . . 
. . . . . . . . . \hfill{38}

\noindent
References  . . . . . . . . . . . . . . . . . . . . . . . . . . . . . . . . . 
. . . . . . . . . . \hfill{41}

\newpage

\section{Introduction}

Quantum Field Theory (QFT) is the theoretical foundation of the elementary
particles physics, including the Standard Model (SM). An experimental success
of the SM gives some general lessons to field theorists. Among them are: (i) 
not just an arbitrary QFT is of importance but only those of them which are
renormalizable, unitary and asymptotically-free gauge theories, (ii) the
crucial role played by various symmetries, including the local (gauge) 
symmetry, internal (rigid) symmetry and supersymmetry, in a `good' QFT, and 
(iii) most of the `good' QFT symmetries, however, have to be broken either 
spontaneously, or quantum-mechanically.

The standard textbook description of quantum gauge theories is often limited
to {\it perturbative} considerations whereas many physical phenomena (e.g., 
confinement) are essentially {\it non}-perturbative. It is usually 
straightforward (although, it may be quite non-trivial~!) to develop the 
quantum perturbation theory in which all the fundamental symmetries are 
manifestly realised. Unfortunately, the perturbative expansion usually does 
not make sense when the field coupling becomes strong. In other words, the 
formal generating functional (path integral) of QFT has to be defined in 
practical terms, and in the past it was actually done in {\it many} ways 
beyond the perturbation theory (e.g., lattice regularization, instantons, 
duality). Because of this reasoning, until recently, it was common to believe 
among most field theorists that {\it any} non-perturbative gauge QFT is not 
well-defined enough, in order to allow one to make definitive predictions and 
non-perturbative calculations 'from the first principles'. 

However, this attitude may have to be revised since the remarkable discovery 
of Seiberg and Witten~\cite{sw} of an {\it exact} non-perturbative solution to 
the low-energy effective action (LEEA) in certain $N=2$ supersymmetric quantum
gauge field theories. Though the non-trivial low-energy solution was found for
the certain class of QFTs having no immediate phenomenological applications, 
it is, nevertheless, of great value since these theories may be a good 
starting point for further symmetry breaking towards the phenomenologically 
applicable QFT models at lower energies, including the SM, while maintaining 
their nice integrability properties at higher energies. 

Among the general lessons of the Seiberg-Witten (SW) exact solution for the 
QFT practitioners are again the same three lessons formulated above in 
relation to the SM (!), this time as regards the non-perturbative story: (i) 
in order to be solvable in the low-energy limit, a QFT has to be the `good' 
one i.e. it should have the $N=2$ extended supersymmetry, (ii) the exact 
symmetries can severely constrain a non-trivial `good' QFT even beyond 
perturbation theory in such a way that a {\it unique} non-perturbative 
solution may exist in the low-energy limit, the SW solution being an example, 
and (iii) many fundamental symmetries (e.g., the non-abelian gauge symmetry 
and supersymmetry) are either already broken in the full non-perturbative QFT,
or they have to be broken further by some dynamical mechanisms, in order to 
make contact with the low-energy phenomenology to be represented by SM. To 
achieve the third goal, one may have to go even beyond the framework of a 
given $N=2$ supersymmetric QFT e.g., by embedding it into a more fundamental 
superstring theory or M-theory in higher dimensions.

Unlike the SM or its minimal ($N=1$) supersymmetric extensions, $N=2$ 
supersymmetric gauge field theories cannot directly serve for phenomenological
applications, partly because an $N=2$ matter can only be defined in {\it 
real}~ representations of the gauge group. Nevertheless, nothing forbids us to
think about an $N=2$ gauge theory as the starting point only, or as the 
intermediate gauge field theory originating from a unified theory (e.g., the
 M-theory) with even higher symmetry or in higher dimensions at larger 
energies. At lower energies, however, the $N=2$ supersymmetric gauge theory is
supposed to be reproduced while $N=2$ supersymmetry should ultimately be 
broken at even lower energies.
\vglue.2in

\subsection{Motivation}

Four-dimensional ($4d$), $N=2$ supersymmetric gauge field theories are not 
integrable, either classically or quantum-mechanically.~\footnote{It is the
{\it self-dual} sector of their Euclidean versions that is integrable in the
classical sense~\cite{mw,kgn}.}  The full quantum effective action $\G$ in 
these theories is highly non-local and intractable. Nevertheless, it can be 
decomposed to a sum of local terms in powers of space-time derivatives or 
momenta to be divided by some dynamically generated scale $\L$ 
(in components), the leading kinetic terms being called the low-energy 
effective action (LEEA). Determining the exact LEEA is a great achievement 
since it provides the information about a non-perturbative spectrum and exact 
static couplings in the full quantum theory at energies below certain scale 
$\L$. Since we are only interested in the $4d$, $N=2$ gauge theories with 
spontaneously broken gauge symmetry via the Higgs mechanism, the effective 
low-energy field theory may include only {\it abelian} massless vector 
particles. All the massive fields (like the charged $W$-bosons) are supposed 
to be integrated out. This very general concept of LEEA is sometimes called 
the {\it Wilsonian} LEEA since it is familiar from statistical mechanics. There
is a difference between the quantum effective action to be defined as the 
generating functional of the one-particle-irreducible (1PI) Green's 
functions or as the Wilsonian effective action, as far as the gauge theories 
with massless particles are concerned. 

$N=2$ supersymmetry severely restricts the form of the LEEA. The very presence
of $N=2$ supersymmetry in the full non-perturbatively defined quantum $N=2$ 
gauge theory follows from the fact that its Witten index~\cite{wind} does not 
vanish, $\D_W=\tr(-1)^F\neq 0$. It just means that $N=2$ supersymmetry cannot 
be dynamically broken.~\footnote{Alternatively, one may prove that the whole 
theory can be consistently defined in a manifestly \newline ${~~~~~}$ 
$N=2$ supersymmetric way, e.g., in $N=2$ superspace.}
 
There are only two basic supermultiplets (modulo classical duality
transformations) in the rigid $N=2$ supersymmetry: an $N=2$ {\it vector}
multiplet and a {\it hypermultiplet}. The $N=2$ vector multiplet components
(in a WZ-gauge) are 
$$\{~ A~,\quad \l^i_{\a}~,\quad V_{\m}~,\quad D^{(ij)} ~\}~,\eqno(1.1)$$ 
where $A$ is a complex Higgs scalar, $\l^i$ is a chiral spinor (`gaugino') 
$SU(2)_A$ doublet, $V_{\m}$ is a real gauge vector field, and $D^{ij}$ is an 
auxiliary scalar $SU(2)_A$ triplet.~\footnote{The internal symmetry $SU(2)_A$ 
here is just the automorphism symmetry of the $N=2$ super-\newline ${~~~~~}$ 
symmetry algebra, that rotates its two spinor supercharges.} Similarly, the 
on-shell physical components of the Fayet-Sohnius (FS)~\cite{fs} version of a 
hypermultiplet are 
$$ {\rm FS}:\qquad \{~ q^i~,\quad \j_{\a}~,\quad\bar{\j}_{\dt{\a}}~ \}~, 
\eqno(1.2)$$ 
where $q^i$ is a complex scalar $SU(2)_A$ doublet, and $\j$ is a Dirac spinor.
There exists another (dual) Howe-Stelle-Townsend (HST) version~\cite{hst} of a 
hypermultiplet, whose on-shell physical components are 
$$ {\rm HST}: \qquad \{~ \o~,\quad \o^{(ij)}~,\quad \c^i_{\a} ~\}~,
\eqno(1.3)$$ 
where $\o$ is a real scalar, $\o^{(ij)}$ is a scalar $SU(2)_A$ triplet, and 
$\c^i$ is a chiral spinor $SU(2)_A$ doublet. The hypermultiplet spinors can
be called `quarks', though it would mean an extra `mirror' particle for each
`true' quark in the $N=2$ super-QCD.

The manifestly supersymmetric formulation of supersymmetric gauge theories
is provided by {\it superspace}~\cite{superspace}. Since superfields are 
reducible representations of supersymmetry, they have to be restricted by 
certain superspace constraints. The standard constraints defining the $N=2$ 
super-Yang-Mills (SYM) theory in the ordinary $N=2$ superspace~\cite{gs} 
essentially amount to the existence of a restricted chiral $N=2$ superfield 
strength $W$, whose leading component is the Higgs field, $\left.W\right|=A$. 
The $N=2$ superfield $W$ contains also the usual Yang-Mills field strength 
$F_{\m\n}(V)$ among its bosonic components, as well as the $SU(2)_A$ auxiliary
triplet $D^{(ij)}$. Since the latter has to be real in the
sence $\Bar{D^{ij}}=\ve_{ik}\ve_{jl}D^{kl}$, it implies ceratain (non-chiral)
$N=2$ superspace constraints on $W$, which are not easy to solve in terms of 
unconstrained $N=2$ superfields in the non-abelian case. The situation is even
more dramatic in the case of the FS hypermultiplet whose off-shell formulation
does not even exist in the ordinary $N=2$ superspace. Though the HST 
hypermultiplet can be defined off-shell in the ordinary $N=2$ superspace, 
where it is sometimes called as an $N=2$ {\it tensor} (or linear) multiplet, 
its self-couplings are very restricted and not universal there. In order to be
coupled to the $N=2$ gauge superfields, the HST hypermultiplet actually has to
be generalised to a reducible (relaxed) version that is highly complicated in 
practice. The most general off-shell formulation of a hypermultiplet is 
however needed e.g., just in order to write down its couplings which may 
appear in the LEEA, in a model-independent way. 

A universal off-shell solution to all $N=2$ supersymmetric field theories was 
proposed in 1984 by Galperin, Ivanov, Kalitzin, Ogievetsky and Sokatchev
~\cite{gikos}. They introduced the so-called $N=2$ {\it harmonic superspace} 
(HSS) by adding the extra bosonic variables (=harmonics) parametrizing 
the sphere $S^2=SU(2)/U(1)$, to the ordinary $N=2$ superspace coordinates. It 
amounts to an introduction of the infinitely 
many auxiliary fields in terms of the ordinary $N=2$ superfields. When using 
the harmonics, one can rewrite the standard $N=2$ superspace constraints to 
another form that may be called a 'zero-curvature representation' in which the
hidden {\it analyticity} structure of the constraints becomes manifest. In this
reformulation, the harmonics play the role of twistors or spectral parameters
that are well-known in the theory of integrable models. As a result, all the 
$N=2$ supersymmetric field theories can be naturally formulated in terms of
{\it unconstrained} so-called {\it analytic} $N=2$ superfields, i.e. fully 
off-shell.~\footnote{See subsect.~3.1. for details about the $N=2$ HSS.} In 
particular, the off-shell FS hypermultiplet is just described by an analytic 
superfield $q^+$ of the $U(1)$ charge $(+1)$, whereas the analytic superspace
measure has the $U(1)$ charge $(-4)$. A generic hypermultiplet Lagrangian in
HSS has to be an analytic function of $q^+$, $\o$ and the harmonics 
$u_i^{\pm}$. In the next subsect.~1.2. we are going to discuss the most 
general form of LEEA, which is dictated by $N=2$ supersymmetry alone. The 
rest of the paper will be devoted to the question how to fix the $N=2$ 
supersymmetric Ansatz for the vector {\it and} hypermultiplet LEEA completely,
when using all the available methods of calculation 
(Fig.~1 [{\it optional~!}\/]).

\begin{figure}
\vglue.1in
\makebox{
\epsfxsize=4in
\epsfbox{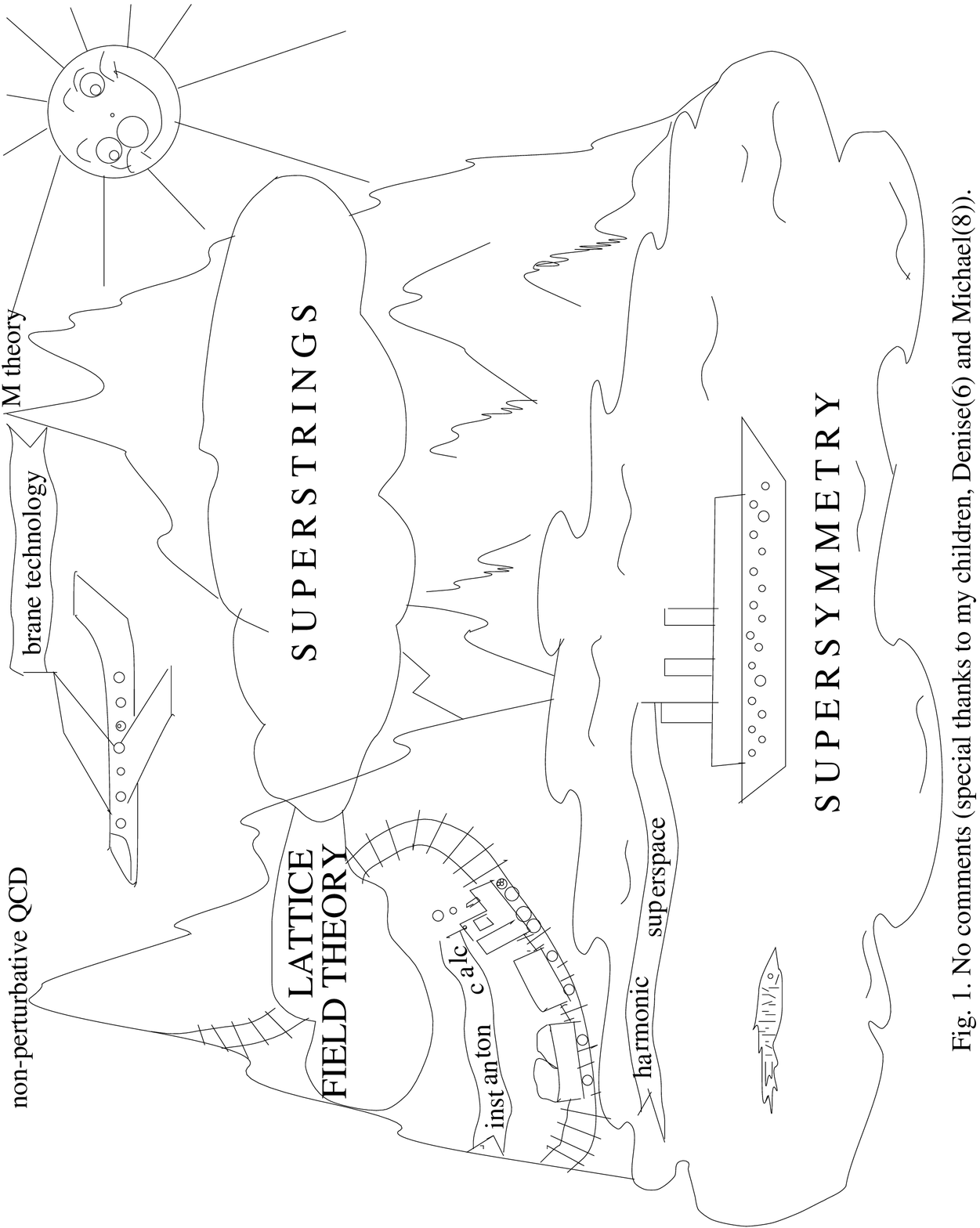}
}
\end{figure}

\subsection{Setup} 

We are now already in a position to formulate the general {\it Ansatz} for the
$N=2$ supersymmetric LEEA. As regards the $N=2$ vector multiplet terms, they
can only be of the form
$$ \G_V[W,\bar{W}]=\int_{\rm chiral} \cf(W) +{\rm h.c.} + \int_{\rm full}
\ch(W,\bar{W}) + \ldots~,\eqno(1.4)$$
where we have used the fact that the abelian $N=2$ superfield strength $W$ is 
an $N=2$ chiral and gauge-invariant superfield. The leading term in eq.~(1.4)
is given by the chiral $N=2$ superspace integral over a {\it holomorphic} 
function $\cf$ of the $W$ that is supposed to be valued in the Cartan 
subalgebra of the gauge group.  The {\it next-to-leading-order} term is given
by the full $N=2$ superspace integral over the real function $\ch$ of $W$ {\it
and}~ $\bar{W}$. The dots in eq.~(1.4) stand for higher-order terms containing
the derivatives of $W$ and $\bar{W}$.

Similarly, the leading non-trivial term in the hypermultiplet LEEA takes the
general form
$$ \G_H[q^+,\sbar{q}{}^+;\o]=\int_{\rm analytic} \ck^{(+4)}(q^+,\sbar{q}{}^+;
\o;u^{\pm}_i)+\ldots~,\eqno(1.5)$$
where $\ck^{(+4)}$ is a function of the FS analytic superfield $q^+$, its 
conjugate $\sbar{q}{}^+$, the HST analytic superfield $\o$ and the harmonics
$u^{\pm}_i$. The action (1.5) is supposed to be added to the kinetic 
hypermultiplet action whose analytic Lagrangian is quadratic in $q^+$ or $\o$,
and of $U(1)$-charge $(+4)$.  A function $\ck$ is called the 
{\it hyper-K\"ahler potential}. When being arbitrarily chosen in eq.~(1.5), 
it automatically leads to the $N=2$ supersymmetric non-linear sigma-model 
(NLSM) with a {\it hyper-K\"ahler} metric, just because of the $N=2$ 
supersymmetry by construction (see an example in subsect.~3.2). 

When being expanded in components, the first term in eq.~(1.4) also leads, 
in particular, to the certain K\"ahler NLSM in the Higgs sector $(A,\bar{A})$.
 The corresponding NLSM K\"ahler potential 
$K_{\cf}(A,\bar{A})$ is dictated by the holomorphic function $\cf$ as $K_{\cf}
={\rm Im}[\bar{A}\cf'(A)]$, so that the function $\cf$ plays the role of a 
potential for this {\it special} NLSM K\"ahler (not hyper-K\"ahler) geometry 
$K_{\cf}(A,\bar{A})$. As regards the hypermultiplet NLSM of eq.~(1.5),
a relation between the hyper-K\"ahler potential $\ck$ and the corresponding 
K\"ahler potential $K_{\ck}$ of the same NLSM is much more involved. Indeed, 
it is easy to see that the hyper-K\"ahler condition on a K\"ahler potential 
amounts to a non-linear (Monge-Ampere) partial differential equation which is 
not easy to solve. It is remarkable that the HSS approach allows one to 
formally get a 'solution' to any hyper-K\"ahler geometry in terms of the 
analytic scalar potential $\ck$. Of course, the real problem is now being 
translated into the precise relation between $\ck$ and the corresponding 
K\"ahler potential (or metric) in components, whose determination amounts to 
solving infinitely many linear differential equations altogether, just in 
order to eliminate the infinite number of the auxiliary fields (see 
subsect.~3.2. for an example). Nevertheless, the hyper-K\"ahler potential 
turns out to be an exteremely useful notion in dealing with the hypermultiplet
LEEA (see sect.~3).

The LEEA gauge-invariant functions $\cf(W)$ and $\ch(W,\bar{W})$ generically 
receive both perturbative and non-perturbative contributions,
$$ \cf= \cf_{\rm per.} + \cf_{\rm inst.}~,\qquad \ch=\ch_{\rm per.} +\ch_{\rm
non-per.}~,\eqno(1.6)$$
while the non-perturbative corrections to the holomorphic function $\cf$ are 
entirely due to instantons. This is an important difference from the (bosonic) 
non-perturbative QCD whose LEEA is dominated by instanton-antiinstanton
contributions.

Unlike the vector LEEA, the exact (charged) hypermultiplet LEEA is essentially
a perturbative one (see sects.~3 and 4), i.e. it does not receive any instanton
corrections,
$$ \ck[q^+] =\ck_{\rm per.}[q^+]~. \eqno(1.7)$$
It is quite remarkable that the perturbative contributions to the leading and
subleading terms in the $N=2$ LEEA entirely come from the {\it one} loop only.
As regards the leading holomorphic contribution, a standard argument goes as 
follows: $N=2$ supersymmetry puts the trace of the energy-momentum tensor 
$T\du{\m}{\m}$ and the axial or chiral anomaly $\pa_{\m} j^{\m}_R$ of the
abelian $R$-symmetry into one $N=2$ supermultiplet. The $T\du{\m}{\m}$ is 
essentially determined by the perturbative renormalization group 
$\b$-function~\footnote{Here and in what
follows $g$ denotes the gauge coupling constant.}, $T\du{\m}{\m}\sim
\b(g)FF$, whereas the one-loop contribution to the chiral anomaly,
$\pa\cdot j_R\sim C_{\rm 1-loop}F{}^*F$, is known to saturate the exact
solution to the Wess-Zumino consistency condition for the same anomaly (e.g.
the one to be obtained from the index theorem). 
Hence, $\b_{\rm per.}(g)=\b_{\rm 1-loop}(g)$ by $N=2$ supersymmetry also. 
Finally, since the $\b_{\rm per.}(g)$ is effectively determined by the second 
derivative of $\cf_{\rm per.}$, one concludes that 
$\cf_{\rm per.}=\cf_{\rm 1-loop}$ too. This simple component argument can be 
extended to a proof in the manifestly $N=2$ supersymmetric way \cite{bko}, 
while the whole chiral perturbative contribution 
$\int_{\rm chiral} \cf_{\rm per.}(W)$ arises in $N=2$ HSS as an anomaly. 
The non-vanishing central charges of the $N=2$ supersymmetry algebra turn out 
to be of crucial importance for the non-vanishing leading holomorphic 
contribution to the gauge LEEA. Its perturbative part takes, therefore, the 
form $\cf_{\rm per.}(W)\sim W^2\log(W^2/\m^2)$, where $\m$ is the 
renormalization group parameter, with the coefficient being fixed by the 
one-loop $\b$-function (see sect.~5). 

The usual strategy in determining the exact LEEA exploits exact symmetries of 
a given $N=2$ quantum gauge theory together with a certain physical input.
As the particular important example of $N=2$ supersymmetric gauge theory, one 
can use the $N=2$ supersymmetric QCD with the gauge group $G_c=SU(N_c)$ and
$N=2$ matter to be described by some number $(N_f)$ of hypermultiplets in the
fundamental representation $\un{N_c}+\un{N_c}^*$ of the gauge group. 
Asymptotic freedom then requires that $N_f<2N_c$.

All possible $N=2$ supersymmetric vacua can be classified as follows:
\begin{itemize}
\item 
{\it Coulomb branch}: $\VEV{q}=0$, while $\VEV{A}\neq 0$; the gauge group 
$G_c$ is broken to its abelian subgroup $U(1)^{{\rm rank}~G_c}$; non-vanishing
'quark' masses are allowed;
\item  {\it Higgs branch}: $\VEV{q}\neq 0$ for some hypermultiplets, while
$\VEV{A}=0$, and all the 'quark' masses vanish; the gauge group $G_c$ is
completely broken;
\item {\it mixed (Coulomb-Higgs) branch}: some $\VEV{q}\neq 0$ {\it and}
$\VEV{A}\neq 0$; it requires $N_c>2$, in particular.
\end{itemize}

In the Coulomb branch, one has to specify the both equations (1.6) and (1.7),
whereas in the Higgs branch only eq.~(1.7) has to be determined. In addition, 
there may be less symmetric vacua when e.g., a non-vanishing Fayet-Iliopoulos 
(FI) term, $\VEV{D^{ij}}=\x^{ij}\neq 0$, is present. $N=2$ supersymmetry may
be (spontaneously) broken by the FI-term too. 
\vglue.2in

\section{Gauge LEEA in the Coulomb branch} 

Seiberg and Witten~\cite{sw} gave a full solution to the holomorphic function
$\cf(W)$ by using certain physical assumptions (about the global structure of
the quantum moduli space $\cm_{\rm qu}$ of vacua) {\it and} electric-magnetic
duality, i.e. not from the first principles. Their main assumption was the
precise value of the Witten index~\footnote{A formal derivation of Witten's
index $\D_W$ from the path integral is plagued with ambiguities.}, $\D_W=2$, 
i.e. exactly two physical singularities in $\cm_{\rm qu}\,$. It is the 
electric-magnetic duality (={\it S-duality}) that was used to connect the weak
and strong coupling regions in $\cm_{\rm qu}\,$.

The Seiberg-Witten solution~\cite{sw} in the simplest
case of the $SU(2)$ gauge group (no fundamental $N=2$ matter) reads
$$ a_D(u) = \fracmm{\sqrt{2}}{\p}\int^u_1\fracmm{dx\sqrt{x-u}}{\sqrt{x^2-1}}~,
\quad
a(u)=\fracmm{\sqrt{2}}{\p}\int^1_{-1}\fracmm{dx\sqrt{x-u}}{\sqrt{x^2-1}}~,
\eqno(2.1)$$
where the renormalization-group independent (Seiberg-Witten) scale $\L^2=1$ 
and, by definition,
$$ a_D=\fracmm{\pa\cf(a)}{da}~.\eqno(2.2)$$
The solution (2.1) is thus written down in the parametric form. Its
holomorphic parameter can be identified with the second Casimir eivenvalue, 
$u=\VEV{\tr A^2}$, that parametrizes $\cm_{\rm qu}\,$. The holomorphic 
function 
$\cf$ can therefore be considered as the one over the quantum moduli space of 
vacua, while the S-duality can be identified with the action of a modular 
group. The monodromies of the multi-valued function $\cf$ around the 
singularities are supplied by the perturbative $\b$-functions, whereas the
whole function $\cf$ is a (unique) solution of the corresponding 
Riemann-Hilbert problem.~\footnote{See refs.~\cite{swrev} for a review.}

In order to make contact with our general discussion in sect.~1, let's consider
the expansion of the SW solution in the semiclassical region, i.e. when 
$\abs{W}\gg\L$,
$$\cf(W)=\fracmm{i}{2\p}W^2\log\fracmm{W^2}{\L^2} +\fracmm{1}{4\p i}W^2
\sum_{m=1}^{\infty} c_m\left(\fracmm{\L^2}{W^2}\right)^{2m}~,\eqno(2.3)$$
where we restored the dependence on $\L$ and written down the interacting 
terms only. It is now obvious that the first term in eq.~(2.3) represents the 
perturbative (one-loop) contribution whereas the rest is just the sum over the
non-perturbative instanton contributions (see subsect.~2.1). It is 
straightforward to calculate the numerical coefficients $\{c_m\}$ from the
explicit solution (2.1) and (2.2)~\cite{klemm}:
$$ \begin{array}{c|cccccc} m & 1 & 2 & 3 & 4 & 5 & \ldots \\
\hline
c_m & 1/2^5 & 5/2^{14} & 3/2^{18} & 1469/2^{31} & 4471/5\cdot 2^{34} & \ldots
\end{array} \eqno(2.4)$$
From the technical point of view, the SW solution is nothing but eq.~(2.4).
It is a challenge for field theorists to reproduce this solution from the
first principles.  
\vglue.2in

\subsection{On the instanton calculations}

The SW solution predicts that the non-perturbative holomorphic contributions 
to the $N=2$ vector gauge LEEA are entirely due to instantons. It is therefore
quite natural to try to reproduce them 'from the first principles', i.e. from 
the path integral approach. The $N=2$ supersymmetric instantons are solutions 
of the classical self-duality equations
$$ F={}^*F~,\qquad i\g^{\m}{D}_{\m}\l=0~,\qquad D^{\m}D_{\m}A
=\[\bar{\l},\l\]~,\eqno(2.5)$$
whose Higgs scalar $A$ approaches a non-vanishing constant $(a)$ at the spacial
infinity so that the whole configuration has a non-vanishing topological charge
$m\in {\bf Z}$. 

From the path-integral point of view, the sum over instantons should be of the
form
$$ \cf_{\rm inst.} =\sum^{\infty}_{m=1}\cf_m~,\quad {\rm where}\quad \cf_m=
\int d \m^{(m)}_{\rm inst.}\,\exp\left[-S_{(m)-{\rm inst.}}\right]~.
\eqno(2.6)$$
Each term $\cf_m$ in this sum can be interpreted as the partition function in 
the multi($m$)-instanton background. The non-trivial measure $d\m^{(m)}_{\rm 
inst.}$ in eq.~(2.6) appears as the result of changing variables from the 
original fields to the collective instanton coordinates in the path integral, 
whereas the $S_{(m)-{\rm inst.}}$ is just the Euclidean action of the $N=2$ 
superinstanton configuration of charge $(m)$. The details about the instanton
calculus can be found e.g., in ref.~\cite{dkm}. One usually assumes that the 
scalar surface term $(\sim \tr\int dS^{\m}A^{\dg}D_{\m}A$) is the only 
relevant term in the action $S_{(m)-{\rm inst.}}$ that contributes. 
In particular, the bosonic and fermionic determinants, that always appear in 
the saddle-point expansion and describe small fluctuations of the fields,
actually cancel in a supersymmetric self-dual gauge background~\cite{adv}. 
Supersymmetry is thus also in charge for the absence of infra-red divergences
present in the determinants.

The functional dependence $\cf_m(a)$ easily follows from the integrated
renormalization group (RG) equation for the one-loop $\b$-function,
$$ \exp\left( -\,\fracmm{8\p^2m}{g^2}\right)
=\left(\fracmm{\L}{a}\right)^{4m}~, \eqno(2.7)$$
and dimensional reasons as follows:
$$ \cf_m(a)=\fracmm{a^2}{4\p i}\left( \fracmm{\L}{a}\right)^{4m}c_m~,
\eqno(2.8)$$
as it should have been expected, up to a numerical coefficient $c_m$. It is
therefore the exact values of the coefficients $\{c_m\}$ that is the issue 
here, as was already noticed above. Their evaluation can thus be reduced 
to the problem of calculating the finite-dimensional multi-instanton measure 
$\{d\m^{(m)}_{\rm inst.}\}$.

A straightforward computation of the measure naively amounts to an explicit
solution of the $N=2$ supersymmetric self-duality equations in terms of the 
collective $N=2$ instanton coordinates for any positive integer instanton 
charge. As is well known, the Yang-Mills self-duality differential equations 
of motion (as well as their supersymmetric counterparts) can be reduced to the
purely {\it algebraic} (though highly non-trivial) set of equations when using
the standard ADHM construction~\cite{adhm}. Unfortunately, an explicit 
solution to the algebraic ADHM equations is known for only $m=1$~\cite{hinst} 
and $m=2$~\cite{osborn}, but it is unknown for $m>2$. Nevertheless, as was 
recently demonstrated by Dorey, Khoze and Mattis~\cite{inmea}, the correct 
multi-instanton measure for any instanton number can be fixed indirectly, by 
imposing $N=2$ supersymmetry and the cluster decomposition requirements alone,
without using the electric-magnetic duality~! It is closed enough to a 
derivation 'from the first principles'. In particular, in the Seiberg-Witten 
model with the $SU(2)$ gauge group considered above, there exists an instanton
solution for $\{c_m\}$ in quadratures~\cite{inmea}. The leading instanton 
corrections for $m=1,2$ do agree with the exact Seiberg-Witten solution 
\cite{iche}. 
\vglue.2in

\subsection{Seiberg-Witten curve}

{} From the mathematical point of view, the Seiberg-Witten exact solution (2.1)
is a solution to the standard Riemann-Hilbert problem of fixing a holomorphic 
multi-valued function $\cf$ by its given monodromy and singularities. The 
number (and nature) of the singularities is the physical input: they are 
identified with the appearance of massless non-perturbative BPS-like physical 
states (dyons) like the famous t'Hooft-Polyakov magnetic monopole. The
monodromies are supplied by perturbative beta-functions and S-duality.

The solution (2.1) can be nicely encoded in terms of the auxiliary 
(Seiberg-Witten) {\it elliptic curve} $\S_{\rm SW}$ defined by the algebraic
equation~\cite{sw}:
$$ \S_{\rm SW}~:\qquad  y^2=(v^2-u)^2-\L^4~.\eqno(2.9)$$
The multi-valued functions $a_D(u)$ and $a(u)$ now appear by integration of a
certain abelian differential $\l$ (of the 3rd kind) over the torus periods 
$\a$ and $\b$ of $\S_{\rm SW}$:
$$ a_D(u)=\oint_{\b}\l~,\qquad a(u)=\oint_{\a} \l~,\quad{\rm where}\quad
\l=v^2\fracmm{dv}{y(v,u)}~.\eqno(2.10)$$

This fundamental relation to the theory of Riemann surfaces can be generalized
further to the other simply-laced gauge groups and $N=2$ super-QCD 
as well~\cite{swgen1,swgen2}. For instance, the solution to the LEEA of the 
pure $N=2$ gauge theory with the gauge group $SU(N_c)$ is encoded in terms of 
the {\it hyperelliptic curve} of genus $(N_c-1)$, whose algebraic equation
reads~\cite{swgen1}
$$ \S_{\rm SW}~:\qquad y^2=W^2_{A_{N_c-1}}(v,\vec{u})-\L^{2N_c}~.\eqno(2.11)$$
The polynomial $W_{A_{N_c-1}}(v,\vec{u})$ introduced above is known in 
mathematics~\cite{agvbook} as the {\it simple singularity} associated with
$A_{N_c-1}\sim SU(N_c)$, or the Landau-Ginzburg super-potential in the $N=2$
supersymmetric $2d$ conformal field theory~\cite{mybook}.  It is given by
$$ W_{A_{N_c-1}}(v,\vec{u})=\sum^{N_c}_{l=1}\left( v - \vec{\l}_l\cdot \vec{a}
\right)=v^{N_c}-\sum^{N_c-2}_{l=0}u_{l+2}(\vec{a})v^{N_c-2-l}~,\eqno(2.12)$$
where $\vec{\l}_l$ are the weights of $SU(N_c)$ in the fundamental 
representation, and $\vec{u}$ are the Casimir eigenvalues, i.e. the Weyl 
group-invariant polynomials in $\vec{a}$ to be constructed by a standard 
(classical) Miura transformation. The simple singularity is the only trace of 
the fundamental non-abelian gauge symmetry in the Coulomb branch.

Adding the (fundamental) $N=2$ matter does not pose a problem in calculating 
the corresponding Seiberg-Witten curve. It reads~\cite{swgen2}
$$ \S_{\rm SW}~:\qquad y^2=W^2_{A_{N_c-1}}(v,\vec{u})-\L^{2N_c-N_f}
\prod^{N_f}_{j=1}(v-m_j)~,\eqno(2.13)$$ 
where $\{m_j\}$ are the bare hypermultiplet masses of $N_f$ hypermultiplets
($N_f<N_c$), in the fundamental representation of the gauge group $SU(N_c)$.

The minimal data $(\S_{\rm SW},\l)$ needed to reproduce the Seiberg-Witten
exact solution can be associated with a certain two-dimensional ($2d$) 
integrable system~\cite{swint}. In addition, the SW potential $\cf$ is a 
solution to the Dijkgraaf-Verlinde-Verlinde-Witten-type~\cite{dvvw} non-linear
differential equations known in the $2d$ (conformal) topological field 
theory~\cite{mmm}:
$$ \cf_i\cf^{-1}_k\cf_j=\cf_j\cf^{-1}_k\cf_i~,\quad{\rm where}\quad
(\cf_i)_{jk}\equiv\fracmm{\pa^3\cf}{\pa a_i\pa a_j \pa a_k}~~.\eqno(2.14)$$
There also exists another non-trivial equation for $\cf$ which is a consequence
of the anomalous (chiral) $N=2$ superconformal Ward identities in 
$4d$~\cite{maw}.

Though the mathematical relevance of the Seiberg-Witten curve is quite clear 
from what was already written above, its geometrical origin and physical 
interpretation are still obscure at this point. It is the issue that can be 
understood in the context of M theory (see sect.~4).
\vglue.2in

\section{Hypermultiplet LEEA in the Coulomb branch}

The previous sect.~2 was entirely devoted to the {\it holomorphic} function 
$\cf$ appearing in the gauge LEEA (1.4) in the Coulomb branch. In this sect.~3
we are going to discuss another {\it analytic} function $\ck$ dictating the
hypermultiplet LEEA (1.5). The function $\ck$ is known as a hyper-K\"ahler 
potential,~\footnote{Any $4d$, globally $N=2$ supersymmetric NLSM with the 
highest physical spin $1/2$ necessarily \newline ${~~~~~}$ has a 
hyper-K\"ahler metric in its kinetic terms~\cite{alfr}.} and 
it plays the role in the hypermultiplet LEEA similar to that of $\cf$ in the 
vector gauge LEEA. Since the very notion of the hyper-K\"ahler potential, in 
fact, requires an introduction of the harmonic superspace (HSS), in the next 
subsect.~3.1 a brief introduction into the $N=2$ HSS is provided 
(see refs.~\cite{gikos,hsf} for more details).
\vglue.2in

\subsection{$N=2$~ harmonic superspace}

The $N=2$ supersymmetric $4d$ field theories can be formulated in a manifestly
$N=2$ supersymmetric way in $N=2$ superspace, in terms of certain constraints.
Unfortunately, the constraints defining a (non-abelian) $N=2$ vector 
multiplet or a hypermultiplet in the ordinary $N=2$ superspace do not have a 
manifestly holomorphic (or analytic) structure, and they 
do not have a simple solution in terms of unconstrained $N=2$ superfields 
which are needed for quantization. The situation is even more dramatic for 
the hypermultiplets whose known off-shell formulations in the ordinary $N=2$
superspace are not universal so that their practical meaning is very limited.  

In the HSS formalism, the standard N=2 superspace~\footnote{Since our 
spacetime is flat we identify the flat $(m=0,1,2,3)$ and curved $(\m=0,1,2,3)$
spacetime \newline ${~~~~~}$ vector indices.} $Z^M
=(x^m,\q^{\a}_i,\bar{\q}^{\dt{\a}i})$, $\a=1,2$, and $i=1,2$, is extended 
by adding the bosonic variables (or `zweibeins') 
$u^{\pm i}$ parameterizing the sphere $S^2\sim SU(2)/U(1)$. By using these
extra variables one can make manifest the hidden analyticity structure of 
all the standard $N=2$ superspace constraints as well as find their
solutions in terms of unconstrained (analytic) superfields. The harmonic
variables have the following fundamental properties:
$$ \left( \begin{array}{c} u^{+i} \\ u^{-i}\end{array}\right) \in SU(2)~,
\quad {\rm so~~that}\quad u^{+i}u^-_i=1~,\quad{\rm and}\quad
 u^{+i}u^+_i=u^{-i}u^-_i=0~.\eqno(3.1)$$
Instead of using an explicit parameterization of the sphere $S^2$, 
it is convenient to deal with functions of zweibeins, that carry a definite 
$U(1)$ charge $q$ to be defined by $q(u^{\pm}_i)=\pm 1$, and use the following 
integration rules~\cite{gikos}:
$$ \int du =1~,\qquad \int du\, u^{+(i_1}\cdots u^{+i_m}u^{-j_1}\cdots
u^{-j_n)}=0~,\quad {\rm when}\quad m+n>0~.\eqno(3.2)$$
It is obvious that any integral over a $U(1)$-charged quantity vanishes.  

The usual complex conjugation does not preserve analyticity (see below).
However, when being combined with another (star) conjugation that only acts on
the $U(1)$ indices as $(u^+_i)^*=u^-_i$ and $(u^-_i)^*=-u^+_i$, it does 
preserve analyticity. One easily finds~\cite{gikos}
$$ \sbar{u^{\pm i}}=-u^{\pm}_i~,\qquad  \sbar{u^{\pm}_i}=u^{\pm i}~.
\eqno(3.3)$$

The covariant derivatives with respect to the zweibeins,  that preserve the 
defining conditions (3.1), are given by 
$$ D^{++}=u^{+i}\fracmm{\pa}{\pa u^{-i}}~,\quad 
D^{--}=u^{-i}\fracmm{\pa}{\pa u^{+i}}~,\quad 
D^{0}=u^{+i}\fracmm{\pa}{\pa u^{+i}}-u^{-i}\fracmm{\pa}{\pa u^{-i}}~.
\eqno(3.4)$$
It is easy to check that they satisfy the $SU(2)$ algebra,
$$\[ D^{++},D^{--}\]=D^0~,\quad \[D^0,D^{\pm\pm}\]=\pm 2D^{\pm\pm}~.
\eqno(3.5)$$
 
The key feature of the $N=2$ HSS is the existence of the so-called
 {\it analytic} subspace parameterized by the coordinates
$$ (\z,u)=\left\{ \begin{array}{c}
x^m_{\rm A}=x^m-2i\q^{(i}\s^m\bar{\q}^{j)}u^+_iu^-_j~,~~
\q^+_{\a}=\q^i_{\a}u^+_i~,~~ \bar{\q}^+_{\dt{\a}}=\bar{\q}^i_{\dt{\a}}u^+_i~,~~
u^{\pm}_i \end{array} \right\}~,\eqno(3.6)$$
which is invariant under $N=2$ supersymmetry, and is closed under the combined 
conjugation of eq.~(3.3)~\cite{gikos}. It allows one to define the 
{\it analytic} 
superfields of any $U(1)$ charge $q$, by the analyticity conditions
$$D^+_{\a}\f^{(q)}=\bar{D}^+_{\dt{\a}}\f^{(q)}=0~,\quad {\rm where}\quad
D^{+}\low{\a}=D^i_{\a}u^+_i \quad {\rm and}\quad
\bar{D}^+_{\dt{\a}}=\bar{D}^i_{\dt{\a}}u^+_i~,\eqno(3.7)$$
and introduce the analytic measure $d\z^{(-4)}du\equiv d^4x_{\rm A}
d^2\q^+d^2\bar{\q}^+du$ of charge $(-4)$, so that the full measure in 
the $N=2$ HSS can be written down as
$$ d^4xd^4\q d^4\bar{\q}du=d\z^{(-4)}du(D^+)^4~,\eqno(3.8)$$
where
$$(D^+)^4=\fracmm{1}{16}(D^+)^2(\bar{D}^+)^2
=\fracmm{1}{16}(D^{+\a}D_{\a}^+)(\bar{D}^{+}_{\dt{\a}}\bar{D}^{+\dt{\a}})~.
\eqno(3.9)$$
In the analytic subspace, the harmonic derivative $D^{++}$ takes the form
$$D^{++}_{analytic} = D^{++}-2i\q^+\s^m\bar{\q}^+\pa_m~,\eqno(3.10)$$ 
it preserves analyticity, and it allows one to integrate by parts. Both the 
original (central) basis and the analytic one can be used on equal footing in 
the HSS. In what follows we omit the subscript {\it analytic}  at the 
covariant derivatives in the analytic basis, in order to simplify the notation.

It is the advantage of the analytic $N=2$ HSS compared to the ordinary $N=2$ 
superspace that both an off-shell $N=2$ vector multiplet and an off-shell
hypermultiplet can be introduced there on equal footing. There exist two 
off-shell hypermultiplet versions in HSS, which are dual to each other.
The so-called {\it Fayet-Sohnius} (FS) hypermultiplet is defined as an 
unconstrained complex analytic superfield $q^+$ of the $U(1)$-charge $(+1)$, 
whereas its dual, called the {\it Howe-Stelle-Townsend} (HST) hypermultiplet, 
is a real unconstrained analytic superfield $\o$ with the vanishing 
$U(1)$-charge.
The on-shell physical components of the FS hypermultiplet comprise an $SU(2)_A$
doublet of complex scalars and a Dirac spinor which is a singlet w.r.t. the
$SU(2)_A$. The on-shell physical components of the HST hypermultiplet comprise
real singlet and triplet of scalars, and a doublet of chiral spinors. The FS
hypermultiplet is natural for describing a charged $N=2$ matter (e.g. in the 
Coulomb branch), whereas the HST hypermultiplet is natural for describing
a neutral $N=2$ matter or the Higgs branch. 
Similarly, an $N=2$ vector multiplet is described by an unconstrained analytic
superfield $V^{++}$ of the $U(1)$-charge $(+2)$. The $V^{++}$ is real in the 
sense $\Bar{V^{++}}^{\,*}=V^{++}$, and it can be naturally introduced as a 
connection to the harmonic derivative $D^{++}$. 

A free FS hypermultiplet action is given by
$$ S[q]=-\int d\z^{(-4)}du\,\sbar{q}{}^+D^{++}q^+~,\eqno(3.11)$$
whereas its minimal coupling to an $N=2$ gauge superfield reads
$$ S[q,V]= -\tr\int d\z^{(-4)}du \,\sbar{q}{}^+(D^{++}+iV^{++})q^+~,
\eqno(3.12)$$
where the both superfields, $q^+$ and $V^{++}$, are now Lie algebra-valued.

It is not difficult to check that the free FS hypermultiplet equations of 
motion, $D^{++}q^+=0$, imply $q^+=q^i(z)u^+_i$ as well as the usual (on-shell)
Fayet-Sohnius constraints~\cite{fs} in the ordinary $N=2$ superspace,
$$D_{\a}^{(i}q^{j)}(z)=D_{\dt{\a}}^{(i}q^{j)}(z)=0~.\eqno(3.13)$$
Similarly, a free action of the HST hypermultiplet is given by
$$S[\o]=-\frac{1}{2}\int d\z^{(-4)}du \,(D^{++}\o)^2~,\eqno(3.14)$$
and it is equivalent on-shell to the standard $N=2$ tensor (or linear)
multiplet (see subsect.~3.3).

The standard Grimm-Sohnius-Wess (GSW) constraints~\cite{gs} defining the 
$N=2$ super-Yang-Mills theory in the ordinary $N=2$ superspace imply 
the existence of a (covariantly) chiral~\footnote{A covariantly-chiral
superfield can be transformed into a chiral superfield by field redefinition.}
and gauge-covariant $N=2$ SYM field strength $W$ satisfying, in addition, the 
reality condition (or the Bianchi `identity') 
$$ {\cal D}^{\a}\low{(i}{\cal D}\low{j)\a}W=\bar{\cal D}_{\dt{\a}(i}
\bar{\cal D}^{\dt{\a}}\low{j)}\bar{W}~.\eqno(3.15)$$
Unlike the $N=1$ SYM theory, an $N=2$ supersymmetric solution to the 
non-abelian $N=2$ SYM constraints in the ordinary $N=2$ superspace is not 
known in an analytic form. It is the $N=2$ HSS reformulation of the $N=2$ SYM 
theory that makes it possible~\cite{gikos}. The exact non-abelian relation 
between the constrained, harmonic-independent superfield strength $W$ and the 
unconstrained analytic (harmonic-dependent) superfield $V^{++}$ is given in 
refs.~\cite{gikos,hsf}, and it is highly non-linear. It is merely its abelian 
version that is needed for calculating the perturbative LEEA in the Coulomb 
branch. The abelian relation is given by
$$ W=\fracmm{1}{4} \{ \bar{\cal D}^+_{\dt{\a}},\bar{\cal D}^{-\dt{\a}}\}
=-\fracmm{1}{4}(\bar{D}^+)^2A^{--}~,\eqno(3.16)$$
where the non-analytic harmonic superfield connection $A^{--}(z,u)$ to the 
derivative $D^{--}$ has been introduced, ${\cal D}^{--}=D^{--}+iA^{--}$.
As a consequence of the $N=2$ HSS abelian constraint 
$\[{\cal D}^{++},{\cal D}^{--}\]={\cal D}^0=D^0$, the connection $A^{--}$ 
satisfies the relation
$$ D^{++}A^{--}=D^{--}V^{++}~,\eqno(3.17)$$
whereas eq.~(3.15) can be rewritten to the form
$$ (D^+)^2W=(\bar{D}^+)^2\bar{W}~.\eqno(3.18)$$

A solution to the $A^{--}$ in terms of the analytic unconstrained superfield
$V^{++}$ easily follows from eq.~(3.17) when using the identity~\cite{hsf}
$$ D^{++}_1(u_1^+u^+_2)^{-2}=D_1^{--}\d^{(2,-2)}(u_1,u_2)~,\eqno(3.19)$$
where we have introduced the harmonic delta-function $\d^{(2,-2)}(u_1,u_2)$ 
and the harmonic distribution $(u_1^+u^+_2)^{-2}$ according to their 
definitions in refs.~\cite{gikos,hsf}, hopefully, in the self-explaining 
notation. One finds~\cite{zupnik} 
$$ A^{--}(z,u)= \int dv \,\fracmm{V^{++}(z,v)}{(u^+v^+)^2}~,\eqno(3.20)$$
and
$$ W(z)=-\fracmm{1}{4}\int du (\bar{D}^-)^2V^{++}(z,u)~,\quad \bar{W}(z)=
-\fracmm{1}{4}\int du (D^-)^2V^{++}(z,u)~,\eqno(3.21)$$
by using the identity 
$$u^+_i=v^+_i(v^-u^+)-v^-_i(u^+v^+)~,\eqno(3.22)$$
which is the obvious consequence of the definitions (3.1). 

The equations of motion are given by the vanishing analytic superfield
$$ (D^+)^4A^{--}(z,u)=0~,\eqno(3.23)$$
while the corresponding action reads~\cite{zupnik}
$$ \eqalign{
S[V]= & \fracmm{1}{4}\int d^4xd^4\theta\, W^2 + {\rm h.c.}=
\fracmm{1}{2}\int d^4xd^4\theta d^4\bar{\theta}du \,V^{++}(z,u)A^{--}(z,u)\cr
= & \fracmm{1}{2}\int d^4xd^4\theta d^4\bar{\theta}du_1du_2\,
\fracmm{V^{++}(z,u_1)V^{++}(z,u_2)}{(u_1^+u^+_2)^2}~.\cr}\eqno(3.24)$$

In a WZ-like gauge, the abelian analytic pre-potential $V^{++}$ amounts 
to the expression~\cite{gikos}
$$\eqalign{
 V^{++}(x_{\rm A},\theta^+,\bar{\theta}^+,u)=&
\bar{\theta}^+\bar{\theta}^+a(x_{\rm A})
+ \bar{a}(x_{\rm A})\theta^+\theta^+ 
-2i\theta^+\s^{m}\bar{\theta}^+V_{m}(x_{\rm A}) \cr
& +\bar{\theta}^+\bar{\theta}^+\theta^{\a +}\j^i_{\a}(x_{\rm A})u^-_i
+\theta^+\theta^+\bar{\theta}^+_{\dt{\a}}\bar{\j}^{\dt{\a}i}(x_{\rm A})u^-_i\cr
&+\theta^+\theta^+\bar{\theta}^+\bar{\theta}^+D^{(ij)}(x_{\rm A})u^-_iu^-_j~,
\cr} \eqno(3.25)$$
where $(a,\j^i_{\a},V_{m},D^{ij})$ are the usual $N=2$ vector multiplet 
components~\cite{gs}.

The (BPS) mass of a hypermultiplet can only come from the central charges 
of the 
$N=2$ SUSY algebra since, otherwise, the number of the massive hypermultiplet 
components has to be increased. The most natural way to introduce central 
charges $(Z,\bar{Z})$ is to identify them with spontaneously broken $U(1)$ 
generators of dimensional reduction from six dimensions via the Scherk-Schwarz
mechanism~\cite{ss}. Being rewritten to six dimensions, 
eq.~(3.10) implies the additional `connection' term in the associated 
four-dimensional harmonic derivative 
$$ {\cal D}^{++}=D^{++}+v^{++}~,\quad {\rm where}\quad 
v^{++}=i(\theta^+\theta^+)\bar{Z}+i(\bar{\theta}^+\bar{\theta}^+)Z~.
\eqno(3.26)$$
Comparing eq.~(3.26) with eqs.~(3.12) and (3.21) clearly shows that the $N=2$ 
central charges can be equivalently treated as a non-trivial $N=2$ gauge
background, with the covariantly constant chiral superfield strength
$$ \VEV{W}=\VEV{a}=Z~,\eqno(3.27)$$
where eq.~(3.25) has been used too. See refs.~\cite{ke,bbi,bb,ikz} for more 
details.
\vglue.2in

\subsection{Taub-NUT metric or KK-monopole}

Since the HSS formulation of hypermultiplets has the manifest off-shell $N=2$
supersymmetry, it is perfectly suitable for discussing possible hypermultiplet 
self-interactions which are highly restricted by $N=2$ supersymmetry. 
Moreover, the manifestly $N=2$ supersymmetric Feynman rules can be derived in 
HSS. The latter can be used to actually calculate the hypermultiplet LEEA (see
below). 

To illustrate the power of HSS, let's consider a single FS hypermultiplet for
simplicity. Its free action in HSS can be rewritten in the pseudo-real 
notation, $q^+_a=(q^+,\sbar{q}{}^+)$, $q^a=\ve^{ab}q_b$, $a=1,2$, as follows:
$$S[q]= -\ha \int_{\rm analytic} q^{a+}{\cal D}^{++}q^+_a~,\eqno(3.28)$$
where the derivative ${\cal D}^{++}$ (in the analytic basis) includes central 
charges in accordance with eq.~(3.26). It is obvious from eq.~(3.28) that the 
action $S[q]$ has the extended internal symmetry given by
$$  SU(2)_A \otimes SU(2)_{\rm extra}~,\eqno(3.29)$$
where the $SU(2)_A$ is the automorphism symmetry of the $N=2$ supersymmetry
algebra~\footnote{It is easy to keep track of the ~$SU(2)_A$~ symmetry in the
~$N=2$~ HSS where this symmetry \newline ${~~~~~}$ amounts to the absence of 
an explicit dependence of a HSS Lagrangian upon the harmonic \newline 
${~~~~~}$ variables $u^{\pm}_i$.} and the 
$SU(2)_{\rm extra}$ acts on the extra indices $(a)$ only. Adding a minimal 
interaction with an abelian $N=2$ vector superfield $V^{++}$ in eq.~(3.28)
obviously breaks the internal symmetry (3.29) down to a subgroup
$$ SU(2)_A\otimes U(1)_e~.\eqno(3.30)$$
It is now easy to see that the {\it only} FS hypermultiplet self-interaction,
that is consistent with the internal symmetry (3.30), is given by the
hyper-K\"ahler potential
$$ \ck^{(+4)}=\fracmm{\l}{2}\left(\sbar{q}{}^+q^+\right)^2~,\eqno(3.31)$$
since it is the only admissible term of the $U(1)$-charge $(+4)$ which can be
added to the FS hypermultiplet action (3.28). We thus get an answer for the
hypermultiplet LEEA in the Coulomb branch almost for free, up to the induced
NLSM coupling constant $\l$.

Similarly, the unique FS hypermultiplet self-interaction in the $N=2$ super-QCD
with $N_c=3$ colors and $N_f$ flavors, and vanishing bare hypermultiplet 
masses, that is consistent with the $SU(N_f)\otimes SU(2)_A\otimes U(1)^2$
internal symmetry, reads
$$ \ck_{\rm QCD}^{(+4)}=\fracmm{\l}{2}\sum^{N_f}_{i,j=1}\left(\sbar{q}{}^{i+}
\cdot q^+_j\right)\left(\sbar{q}{}^{j+}\cdot q^+_i\right)~,\eqno(3.32)$$
where the dots stand for contractions of color indices.

\begin{figure}
\vglue.1in
\makebox{
\epsfxsize=4in
\epsfbox{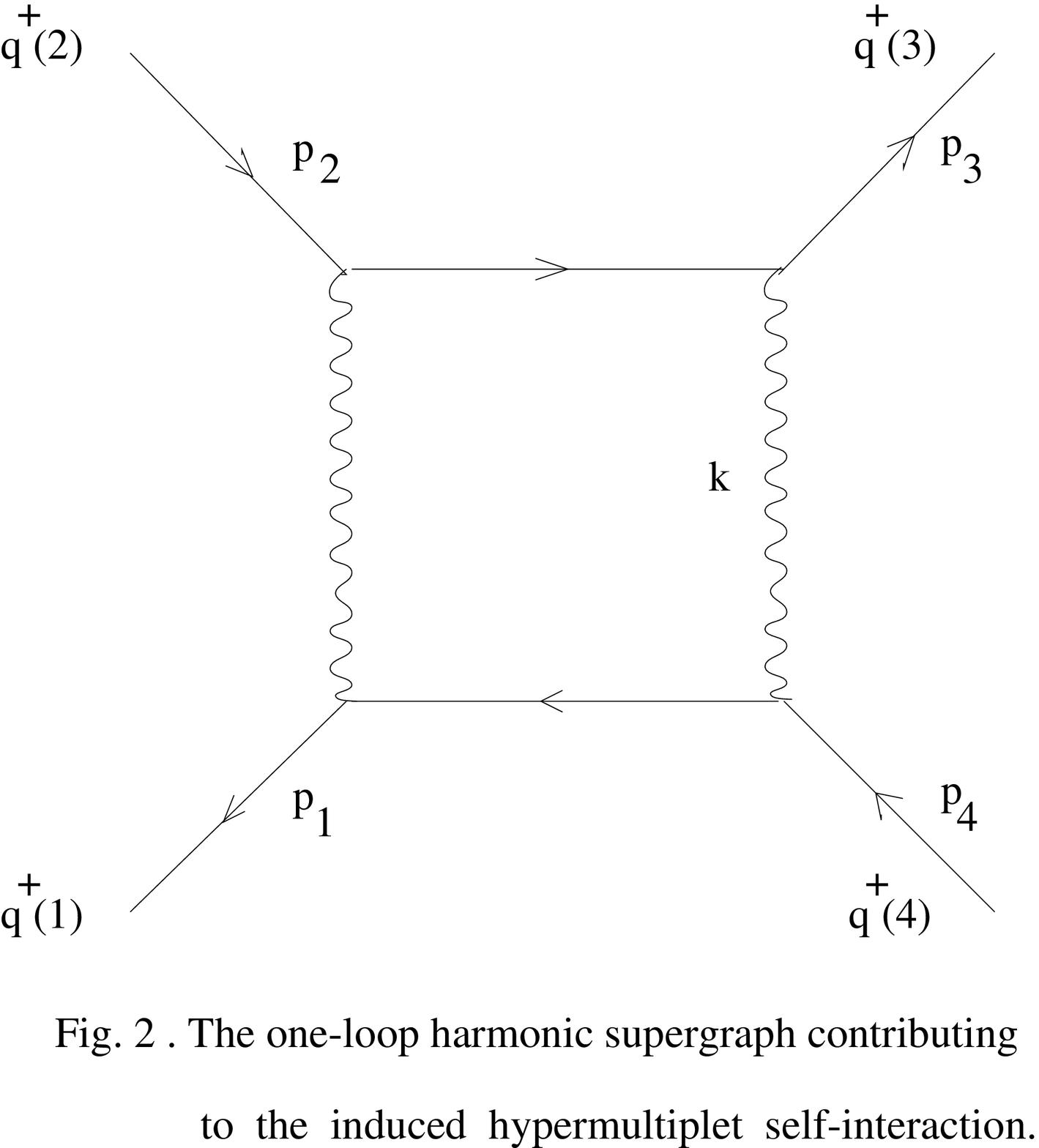}
}
\end{figure}

The induced coupling constant $\l$ in eq.~(3.31) is entirely determined by
the one-loop HSS graph shown in Fig.~2. Since the result vanishes $(\l=0)$
in the absence of central charges,~\footnote{The same conclusion also follows
from the $N=1$ superspace calculations~\cite{gru}.} let's assume that
$Z=a\neq 0$, i.e. we are in the Coulomb branch. The free HSS actions of an
$N=2$ vector multiplet and a hypermultiplet given above are enough to compute
the corresponding $N=2$ superpropagators. The $N=2$ vector multiplet action
takes the particularly simple form in the $N=2$ super-Feynman gauge (there are
no central charges for the $N=2$ vector multiplet), 
$$ S[V]_{\rm Feynman}=\ha \int_{\rm analytic} V^{++}\Box V^{++}~,\eqno(3.33)$$
so that the corresponding analytic HSS propagator (the wave lines in Fig.~2) 
reads 
$$ i\VEV{ V^{++}(1)V^{++}(2)}=\fracmm{1}{\Box_1}(D_1^+)^4\d^{12}(Z_1-Z_2)
\d^{(-2,2)}(u_1,u_2)~,\eqno(3.34)$$
where the harmonic delta-function $\d^{(-2,2)}(u_1,u_2)$ has been 
introduced~\cite{hsf}. The FS hypermultiplet HSS propagator (solid lines in
Fig.~2) with non-vanishing central charges is more 
complicated~\cite{zupnik,ikz}:
$$ i\VEV{q^+(1)q^+(2)}=\fracmm{-1}{\Box_1+a\bar{a}}
\fracmm{(D^+_1)^4(D^+_2)^4}{(u^+_1u_2^+)^3}e^{\t_3[v(2)-v(1)]}\d^{12}
(Z_1-Z_2)~,\eqno(3.35)$$
where $v$ is the so-called 'bridge' satisfying the equation 
${\cal D}^{++}e^v=0$. One easily finds that
$$ iv =-a(\bar{\theta}^+\bar{\theta}^-)-\bar{a}(\theta^+\theta^-)~.\eqno(3.36)
$$
The rest of the $N=2$ HSS Feynman rules is very similar to that of the 
ordinary $(N=0)$ Quantum Electrodynamics (QED). 

A calculation of the HSS graph in Fig.~2 is now straightforward, while the 
calculational details are given in ref.~\cite{ikz}. One finds the predicted
form of the induced hyper-K\"ahler potential as in eq.~(3.31) indeed, with
the induced NLSM coupling constant given by
$$ \l=\fracmm{g^4}{\p^2}\left[ \fracmm{1}{m^2}\ln\left( 1+\fracmm{m^2}{\L^2}
\right) -\fracmm{1}{\L^2+m^2}\right]~,\eqno(3.37)$$
where $g$ is the gauge coupling constant, $m^2=\abs{a}^2$ is the hypermultiplet
BPS mass, and $\L$ is the IR-cutoff parameter. Note that $\l\to 0$ when the
central charge $a\to 0$.

Yet another technical problem is how to decode the HSS result (3.31) in the 
conventional component form. In other words, one still has to find an explicit
hyper-K\"ahler metric that corresponds to the hyper-K\"ahler potential (3.31).
The general procedure of getting the component form of the bosonic NLSM from a
hypermultiplet self-interaction in HSS consists of the following steps:
\begin{itemize}
\item expand the equations of motion in the Grassmann (anticommuting) 
coordinates, and ignore all the fermionic field components,
\item solve the kinematical linear differential equations for all the 
auxiliary fields, thus eliminating the infinite tower of them in the harmonic 
expansion of the hypermultiplet HSS analytic superfields~;
\item substitute the solution back into the HSS hypermultiplet action, and
integrate over all the anitcommuting and harmonic HSS coordinates.
\end{itemize}

Of course, it is not always possible to actually perform this procedure. For
instance, just the second step above would amount to solving infinitely many 
linear differential equations altogether. However, just in the case of 
eq.~(3.31), the explicit solution exists~\cite{giost,ikz}. When using the
parametrization
$$ \left. q^+\right|_{\theta=0}=f^i(x)u^+_i\exp\left[ \l f^{(j}(x)\bar{f}^{k)}
(x)u^+_j u^-_k\right]~,\eqno(3.38)$$
one finds the $4d$ bosonic NLSM action
$$ S_{\rm NLSM}=\int d^4x\,\left\{ g_{ij}\pa_mf^i\pa^mf^j +\bar{g}^{ij}
\pa_m\bar{f}_i\pa^m\bar{f}_j +h\ud{i}{j}\pa_mf^j\pa^m\bar{f}_i -V(f)
\right\}~,\eqno(3.39)$$
whose metric is given by~\cite{giost}
$$ g_{ij}= \fracmm{ \l(2+\l f\bar{f}}{4(1+\l f \bar{f})}\bar{f}_i\bar{f}_j~,
\quad \bar{g}^{ij}=\fracmm{\l(2+\l f\bar{f}}{4(1+\l f\bar{f})}f^if^j~,$$
$$h\ud{i}{j}=\d\ud{i}{j}(1+\l f\bar{f})-\fracmm{\l(2+\l f\bar{f})}{2(1+
\l f\bar{f}}f^i\bar{f}_j~,\quad f\bar{f}\equiv f^i\bar{f}_i~,\eqno(3.40)$$
whereas the induced scalar potential reads~\cite{ikz}
$$V(f) =\abs{Z}^2\fracmm{f\bar{f}}{1+\l f\bar{f}}~~~.\eqno(3.41)$$

In the form (3.40) the induced metric is apparently free from any 
singularities. It is generically non-trivial to compare the induced NLSM
metric with any standard hyper-K\"ahler metric since the NLSM metrics 
themselves are defined modulo field redefinitions, i.e. modulo four-dimensional
 diffeomorphisms in the case under considerarion. Fortunately, it is known how
to transform the metric (3.40) to the standard Taub-NUT form~:
$$ ds^2=\fracmm{r+M}{2(r-M)}dr^2+\ha(r^2-M^2)(d\vartheta^2+\sin^2\vartheta
d\varphi^2)+2M^2\left( \fracmm{r-M}{r+M}\right)
(d\j+\cos\vartheta d\varphi)^2~, \eqno(3.42)$$
by using the following change of variables~\cite{giost}:
$$f^1= \sqrt{2M(r-M)}\cos\fracmm{\vartheta}{2}\exp\fracmm{i}{2}(\j+\varphi)~,
$$
$$f^2= \sqrt{2M(r-M)}\sin\fracmm{\vartheta}{2}\exp\fracmm{i}{2}(\j-\varphi)~,
 \eqno(3.43)$$
$$ f\bar{f}=2M(r-M)~,\qquad r\geq M=\fracmm{1}{2\sqrt{\l}}~,$$
where $M=\ha\l^{-1/2}\sim g^{-2}$ is the mass of the Taub-NUT instanton, also 
known as the KK-instanton~\cite{town}.

Therefore, the induced metric of the hypermultiplet LEEA in the Coulomb branch
is generated in one loop, and it is given by the Taub-NUT or 
its higher-dimensional generalizations. The non-trivial scalar potential is 
also generated by quantum corrections, with the BPS mass being unrenormalized 
as it should.
\vglue.2in

\subsection{Duality transformation and $N=2$ tensor multiplet}

There exists an interesting connection between the FS hypermultiplet Taub-NUT
self-interaction in the $N=2$ {\it harmonic} superspace and the $N=2$ tensor
(or linear) multiplet self-interaction in the {\it ordinary} $N=2$ superspace.
Namely, the $N=2$ supersymmetric Taub-NUT NLSM is equivalent to a sum of the
naive (quadratic in the fields and non-conformal) and {\it improved} 
(non-polynomial in the fields and $N=2$ superconformally invariant) actions
for the $N=2$ tensor multiplet in the ordinary $N=2$ superspace~!

It is worth mentioning here that the $N=2$ {\it tensor} multiplet in the 
ordinary $N=2$ superspace is defined by the constraints
$$ D\low{\a}{}^{(i}L^{ik)}(Z)=\bar{D}_{\dt{\a}}{}^{(i}L^{jk)}(Z)=0~,\eqno(3.44)$$
and the reality condition
$$ \Bar{L^{ij}}=\ve_{ik}\ve_{jl}L^{kl}~.\eqno(3.45)$$
Unlike the FS hypermultiplet in the ordinary $N=2$ superspace, the constraints
(3.44) are off-shell, i.e. they do not imply the equations of motion for the
components of the $N=2$ tensor multiplet. The $N=2$ tensor multiplet itself 
can be identified as a restricted HST hypermultiplet (i.e. as an analytic $\o$ 
superfield subject to extra off-shell constraints), while its $N=2$
supersymmetric self-interactions are a subclass of those for $\o$~\cite{gio}.
The $N=2$ tensor multiplet has $8_{\rm B}\oplus 8_{\rm F}$ off-shell 
components:
$$ \vec{L},\quad \z^i_{\a}~,\quad B~,\quad {E'}_{m}=\ha\ve_{mnpq}\pa_nE_{pq}~,
\eqno(3.46)$$
where $\vec{L}$ is the scalar $SU(2)_A$ triplet, $\vec{L}=\tr(\vec{\t}L)$ and
$\vec{\t}$ are Pauli matrices, $\z^i_{\a}$ is a chiral spinor doublet, $B$ is a
complex auxiliary scalar, and $E_{mn}$ is a gauge antisymmetric tensor whose
field strength is ${E'}_m$.

Let's start with our induced hypermultiplet LEEA
$$ S[q^+]_{\rm Taub-NUT} = \int_{\rm analytic} \left[ \sbar{q}{}^+D^{++}q^+
+ \fracmm{\l}{2}(q^+)^2(\sbar{q}{}^+)^2\right]~,\eqno(3.47)$$
and make the following substitution of the HSS superfield variables~\cite{gio}:
$$ \sqrt{\l}q^+=-i\left(2u^+_1+ig^{++}u^-_1\right)e^{-i\tilde{\o}/2}~,
\quad \sqrt{\l}\sbar{q}{}^+=i\left( 2u^+_2-ig^{++}u^-_2\right)e^{i\tilde{\o}
/2}~,\eqno(3.48)$$
where
$$ g^{++}(l,u)\equiv \fracmm{2(l^{++}-2iu_1^+u_2^+)}{1+\sqrt{1-4u^+_1u^+_2
u^-_1u^-_2 -2il^{++}u_1^-u_2^-}}~,\eqno(3.49)$$
and $(l^{++},\o)$ are the new dimensionless analytic superfieds. It is not
difficult to check that eqs.~(3.48) and (3.49) imply, in particular, that
$$ \l \sbar{q}{}^+q^+=2il^{++}~,\eqno(3.50)$$
whereas the action (3.47) takes the form (after the rescaling $l^{++}\equiv
\sqrt{\l}L^{++}$ and $\tilde{\o}=\sqrt{\l}\o$):
$$ S[L^{++};\o]\low{\rm Taub-NUT}=
S\low{\rm free}[L^{++};\o] + S\low{\rm impr.}[L^{++};\o]~,\eqno(3.51)$$
where 
$$S\low{\rm free}[L^{++};\o]= 
\ha \int_{\rm analytic}\left[ (L^{++})^2+\o D^{++}
L^{++}\right]~,\eqno(3.52)$$
and
$$ S\low{\rm impr.}[L^{++};\o]= 
\fracmm{1}{2\l}\int_{\rm analytic} \left[g^{++}
(L;u)\right]^2~.\eqno(3.53)$$

The action (3.51) or (3.52) contains $\o$ as a Lagrange multiplier. On the
one hand, varying it w.r.t. $\o$ yields the constraint
$$ D^{++}L^{++}=0~,\eqno(3.54)$$
that, in its turn, implies $L^{++}=u^+_iu^+_jL^{ij}(Z)$ {\it and } 
eq.~(3.44). Therefore, the actions (3.51) and (3.52) describe an $N=2$ tensor 
multiplet in the $N=2$ HSS. On the other hand, one can vary the action (3.52) 
w.r.t. $L^{++}$ first. One finds  that
$$ L^{++}=D^{++}\o~.\eqno(3.55)$$
Hence, $L^{++}$ can be removed altogether in favor of $\o$. It is just a
manifestation of the existing classical {\it duality} between the FS 
hypermultiplet $q^+$ and the HST hypermultiplet $\o$ in the $N=2$ HSS. 

The action (3.53) describes the so-called {\it improved} $N=2$ tensor 
multiplet~\cite{impr}. It can be shown that it is fully invariant under the
rigid $N=2$ superconformal symmetry, while the associated hyper-K\"ahler
metric is equivalent to the flat metric up to a $4d$ diffeomorphism 
\cite{impr}. However, the sum of the actions (3.52) and (3.53) describes an 
interacting theory, and it is just the Taub-NUT or the KK- monopole.

Because of this connection between certain $N=2$ supermultiplets and their 
self-interactions in the HSS, it should not be very surprising that the
Taub-NUT self-interaction can also be reformulated in the {\it ordinary} $N=2$
superspace in terms of the $N=2$ tensor multiplet alone, just as a sum of its
naive and improved actions. The most elegant formulation of the latter exists
in the {\it projective} $N=2$ superspace~\cite{klr,myrev} in which the harmonic
variables are replaced by a single complex projective variable $\x\in CP(1)$.
Unlike the $N=2$ HSS, the projective $N=2$ superspace does {\it not} introduce
any extra auxiliary fields beyond those already present in the off-shell $N=2$
tensor multiplet. The starting point now are the defining constraints (3.44)
for the $N=2$ tensor multiplet in the ordinary $N=2$ superspace. It is not
difficult to check that they imply (see ref.~\cite{myrev} for more details and
generalizations)
$$ \de_{\a}G\equiv (D^1_{\a}+\x D^2_{\a})G=0~,\qquad \D_{\dt{\a}}G\equiv 
(\bar{D}^1_{\dt{\a}}+\x\bar{D}^2_{\dt{\a}})G=0~,\eqno(3.56)$$
for {\it any} function $G(Q(\x),\x)$ that is a function of
$Q(\x)\equiv \x_i\x_j L^{ij}(Z)$ and $\x_i\equiv (1,\x)$ only.

It follows that we can build an $N=2$ superinvariant just by integrating $G$
over the rest of the $N=2$ superspace coordinates in the directions which are
'orthogonal' to those in eq.~(3.56), namely,
$$ S_{\rm inv.}[L]= \int d^4x\fracmm{1}{2\p i}\oint_C d\x\,\tilde{\de}^2
\tilde{\D}^2G(Q,\x)~,\eqno(3.57)$$
where we have introduced the new derivatives
$$ \tilde{\de}_{\a}=\x D^1_{\a}-D^2_{\a}~,\quad \tilde{\D}_{\dt{\a}}=\x
\bar{D}^1_{\dt{\a}}-\bar{D}^2_{\dt{\a}}~.\eqno(3.58)$$

The choice of the function $G(Q,\x)$ and the contour $C$ in the complex 
$\x$-plane, which yields the Taub-NUT self-interaction in eq.~(3.57), is
given by~\cite{klr,myrev}
$$ S\low{\rm Taub-NUT}[L]= 
\int d^4x\, \tilde{\de}^2\tilde{\D}^2\fracmm{1}{2\p i}
\left\{ \oint_{C_1} d\x\,\fracmm{Q^2}{2\x} +\fracmm{1}{\sqrt{\l}}\oint_{C_2}
d\x\, Q\ln(\sqrt{\l} Q)\right\}~,\eqno(3.59)$$
where the contour $C_1$ goes around the origin, whereas the contour $C_2$
encircles the roots of the quadratic equation $Q(\x)=0$ in the complex
$\x$-plane.

Finally, one may wonder, in which sense an $N=2$ tensor multiplet action
describes a $4d$, $N=2$ supersymmetric NLSM with the highest physical spin 
$1/2$, because of the apparent presence of the gauge antisymmetric tensor 
$E_{mn}$ among the $N=2$ tensor multiplet components --- see eq.~(3.46). A 
detailed investigation of the component action, that follows from the 
superspace action (3.59), shows that the tensor $E_{mn}$ and its field 
strength ${E'}_m$ enter the action only in the combination 
$$ \left( 1+ \fracmm{1}{\l\vec{L}^2}\right)
({E'}_m)^2+\ha\ve_{mnpq}E_{pq}F_{mn}(L)~,\eqno(3.60)$$
where the tensor
$$ F_{mn}(L)\equiv \left(\pa_m\vec{L}\times\pa_n\vec{L}\right)\cdot
\fracmm{\vec{L}}{\abs{\vec{L}}^3}\eqno(3.61)$$
is formally identical to the electromagnetic field strength of a magnetic 
monopole. Therefore, there exists a vector potential $A_m$ such that 
$F_{mn}(L)=\pa_mA_n -\pa_nA_m$. An explicit magnetic monopole solution for the
locally defined potential $A_m(\vec{L})$ cannot be 'rotationally' invariant 
w.r.t. the $SO(3)\sim SU(2)_A/Z_2$ symmetry, but it can be written down as a 
function of the $SO(2)$-irreducible $L^{ij}$-components to be defined by 
$L^{ij}=\d^{ij}S+P^{(ij)}_{\rm traceless}$. After integrating
by parts and introducing a Lagrange multiplier $V$ as
$$ {}*EF={}^*EdA\to -d{}^*EA=-{E'}_mA_m\to -E_mA_m-E_m\pa_mV~,\eqno(3.62)$$
we can integrate out the full vector $E_m$. It leaves us with the bosonic NLSM
action in terms of the four real scalars $(S,P^{(ij)}_{\rm traceless},V)$ only.
\vglue.2in

\section{Brane technology}

The exact solutions to the LEEA of $4d$, $N=2$ supersymmetric gauge theories 
can be interpreted in a nice geometrical way, when considering these quantum
supersymmetric field theories in the {\it common world-volume} of the (type IIA
superstring or M-theory) {\it branes}~\cite{five,hw,wi}. 

The relevant brane configuration in the type IIA piture,
in ten dimensions: 
$$R^{1+9}\sim(x^0,x^1,x^2,x^3,x^4,x^5,x^6,x^7,x^8,x^9)~,$$
is schematically pictured in Fig.~3. It consists of two solitonic (NS) 
5-branes carrying no RR-charges, $N_c$ Dirichlet-4-branes stretching between 
the 5-branes, and $N_f$ Dirichlet-6-branes~\cite{wi}.

\begin{figure}
\vglue.1in
\makebox{
\epsfxsize=4in
\epsfbox{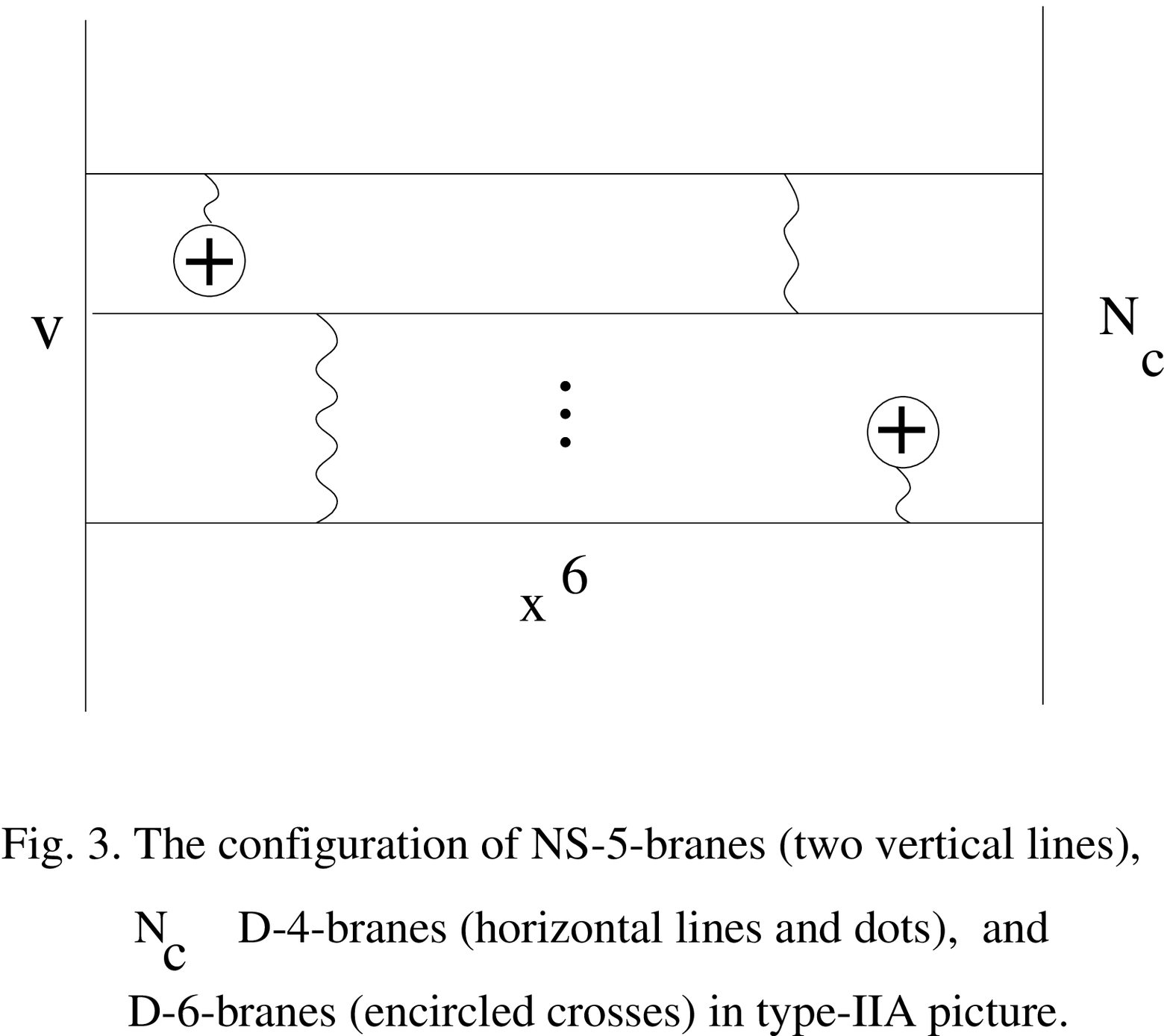}
}
\end{figure}

The two parallel 5-branes are located at $\vec{w}=(x^7,x^8,x^9)=0$ and
(classically) fixed $x^6$ values. Their world-volumes are $R^{1+3}\otimes
\S_0$, where $R^{1+3}$ is the effective (uncompactified) $4d$ spacetime
parametrized by $(x^0,x^1,x^2,x^3)$, and $\S_0$ is a (singular) Riemann 
surface of genus $N_c-1$, parametrized by $(x^4,x^5)$; $v\equiv x^4+ix^5$.

The D-4-branes are also parallel to each other, but are orthogonal to the
5-branes. Their world-volumes are parametrized by $(x^0,x^1,x^2,x^3)\sim
R^{1+3}$ and $x^6$.

The D-6-branes are orthogonal to both 5-branes and D-4-branes, 
they are located at fixed values of $(x^4,x^5,x^6)$, and their 
wolrd-volumes are parametrized by $(x^0,x^1,x^2,x^3)\sim R^{1+3}$ and
$\vec{w}\in B$, where $B$ is the internal space of the D-6-branes.

An $SU(N_c)$  gauge $N=2$ vector multiplet in $R^{1+3}$  appears as the
 massless mode of an open $(4-4)$ string stretching between the D-4-branes,
whereas $N_f$ hypermultiplets in $R^{1+3}$ come from $N_f$ open $(6-4)$
strings connecting the D-6-branes to the D-4-branes. Their BPS masses are
fixed by the distance (in $x^{4,5}$) between the D-6-branes and D-4-branes.

The effective $4d$ coupling constant $g$ is determined by the distance 
between the 5-branes,
$$ \fracmm{1}{g^2}=\fracmm{x^6_1-x^6_2}{2\l}~,\eqno(4.1)$$
where $\l$ is the type-IIA superstring coupling constant. 

The whole brane configuration schematically pictured in Fig.~3 is supposed
to be 'blown up' in order to accommodate the non-perturbative physics. Still,
in this type-IIA picture, it inevitably suffers from quantum singularities
to be associated with the intersections of the NS 5-branes with the 
D-4-branes. These singularities cannot be described semi-classically so 
that one needs yet another resolution~\cite{wi} that is going to be provided 
by reinterpreting the brane configuration of Fig.~3 in M theory (see the next 
subsect.~4.1).

Among the basic properties of the brane configuration under consideration,
let's emphasize the following ones:
\begin{itemize}
\item macroscopically, it is $(1+3)$-dimensional;
\item it is a BPS-like solution to the eleven-dimensional supergravity;
\item it is invariant under $\ha\cdot\ha\cdot 32=8$ supercharges, since
the type-IIA superstring has $32$ supercharges, while a half of them is
conserved by the 5-branes, whereas a half of the remaining $16$ supercharges 
is still conserved by the D-4-branes; being orthogonal to the 5-branes and 
D-4-branes, the D-6-branes do not break any more supercharges; the eight 
supercharges in four dimensions imply the $N=8/4=2$ extended supersymmetry in 
the effective spacetime $R^{1+3}$;
\item the ten-dimensional Lorentz group is spontaneously broken to
$$ SO(1,3)\otimes SU(2)_A\otimes U(1)_{\rm c.c.} \eqno(4.2)$$
\end{itemize}
These are just the properties that actually determine the brane 
configuration of Fig.~3, driven by a desire to have the $N=2$ extended
supersymmetry in the effective $4d$ quantum field theory in the common brane
world-volume as the effective spacetime $R^{1+3}$.
\vglue.2in

\subsection{M-theory resolution}

It is obvious from eq.~(4.1) that one can keep the effective $4d$ gauge 
coupling constant $g$ fixed while {\it increasing} the distance $L=x^6_1-
x^6_2$ between the two 5-branes as well as the type-IIA superstring coupling
constant $\l$. At strong coupling, the type-IIA superstring becomes the 
M-theory~\cite{wm}: one extra dimension $(x^{10})$ to be represented by a
circle $S^1$ of radius $R\sim \l^{2/3}$ shows up, as well as the 
non-perturbative $U(1)_M$ gauge symmetry appears. The latter is associated 
with the $S^1$-rotations.

As a result, the low-energy description of M-theory and its branes suffices
for a geometrical interpretation of the exact solutions to the 
four-dimensional LEEA of the effective $N=2$ supersymmetric gauge theories in 
the M-theory brane world-wolume, just because all the relevant distances in 
the non-perturbative eleven-dimensional brane configuration become {\it large}
while no singularity appears unlike that in the type-IIA picture. 
In particular, the D-4-branes and NS-5-branes in the type-IIA picture are now 
replaced by a {\it single} and {\it smooth} M-theory 5-brane whose
world-volume is given by $R^{1+3}\otimes\S$, where $\S$ is a genus $(N_c-1)$
Riemann surface {\it holomorphically} embedded into a four-dimensional 
hyper-K\"ahler manifold $Q$.~\footnote{The hyper-K\"ahler geometry of $Q$ is, 
in fact, required by $N=2$ supersymmetry in the effective \newline ${~~~~~}$ 
(macroscopic) spacetime $R^{1+3}$.}  The manifold $Q$ is topologically a 
bundle 
$Q\sim B\times S^1$ parametrized by the coordinates $x^4,x^5,x^6$ and $x^{10}$,
 whose base $B$ is the hiden part of the D-6-brane in the type-IIA picture 
and whose fiber $S^1$ is the hidden eleventh dimension of M theory~\cite{town}.

The origin of the abelian gauge fields in the Coulomb branch of the $4d$ gauge
theory also becomes more transparent from the M-theory point of 
view~\cite{five}. As is well-known, there exists a two-form in the M-theory
5-brane world-volume, whose field strength (3-form) $T$ is self-dual (see 
e.g., ref.~\cite{five}).~\footnote{The eight conserved supercharges then imply 
the existence of a six-dimensional self-dual massless \newline ${~~~~~}$ 
(tensor) supermultiplet in the 5-brane world-volume.}  Since the M-theory 
5-brane is wrapped over the Riemann surface, i.e. its  world-volume is 
locally  a product $R^{1+3}\otimes\S_{N_c-1}$, one can decompose the self-dual
 3-form $T$ as
$$ T = F\wedge \o + {}^*F\wedge {}^*\o~,\eqno(4.3)$$
where $F$ is a two-form on $R^{1+3}$, and $\o$ is a one-form on the Riemann
surface $\S_{N_c-1}$ of genus $N_c-1$. The equations of motion $dT=0$ now
imply 
$$ dF=d{}^*F=0~,\eqno(4.4)$$
and
$$ d\o=d{}^*\o=0~.\eqno(4.5)$$
Eq.~(4.5) means that the one-form $\o$ is harmonic on $\S_{N_c-1}$. Since the
number of independent harmonic one-forms on a Riemann surface exactly equals
to its genus~\cite{fkra}, one also has $(N_c-1)$ two-forms $F$, while each
of them satisfies eq.~(4.4). But eq.~(4.4) is nothing but the Maxwell 
equations for an electro-magnetic field strength $F$. It explains the origin
of the gauge group $U(1)^{N_c-1}$ in the Coulomb branch of the effective $4d$
gauge theory. 

The geometrical interpretation of the $N=2$ {\it gauge} LEEA in the Coulomb 
branch of the effective $4d$ gauge field theory is provided by the 
identification~\cite{five,wi}
$$ \S_{N_c-1} = \S_{SW}~.\eqno(4.6)$$
It order to understand the {\it hypermultiplet} LEEA in a similar way, one
first notices that the D-6-branes~\footnote{The D-6-branes and their dual
D-0-branes in the type-IIA picture are of Kaluza-Klein origin.} are {\it
magnetically} charged w.r.t. the non-perturbative $U(1)_M$ symmetry. Hence,
the fiber $S^1$ of $Q$ has to be non-trivial, i.e. of non-vanishing magnetic 
charge (or the first Chern class) $m\neq 0$. When taking into account the 
$U(2)$ isometry of the hyper-K\"ahler manifold $Q_{(m)}$, one concludes that 
$Q_{(m)}$ is necessarily a multi-Taub-NUT space or a (multi)-KK monopole,
just because it is the {\it only} space among the asymptotically locally flat
(ALF) spaces $B\otimes S^1$, whose fibration $S^1$ is non-trivial. In 
particular, when $m=1$, one gets the Taub-NUT space whose metric was already 
described previously in subsect.~3.2.
\vglue.2in

\subsection{Relation to the HSS results and S duality}

The relation between the HSS results of subsect.~3.2 and the brane technology
of subsect.~4.1 towards the hypermultiplet LEEA (in fact, their equivalence)
is provided by the {\it S-duality} in field theory (Fig.~4).

\begin{figure}
\vglue.1in
\makebox{
\epsfxsize=4in
\epsfbox{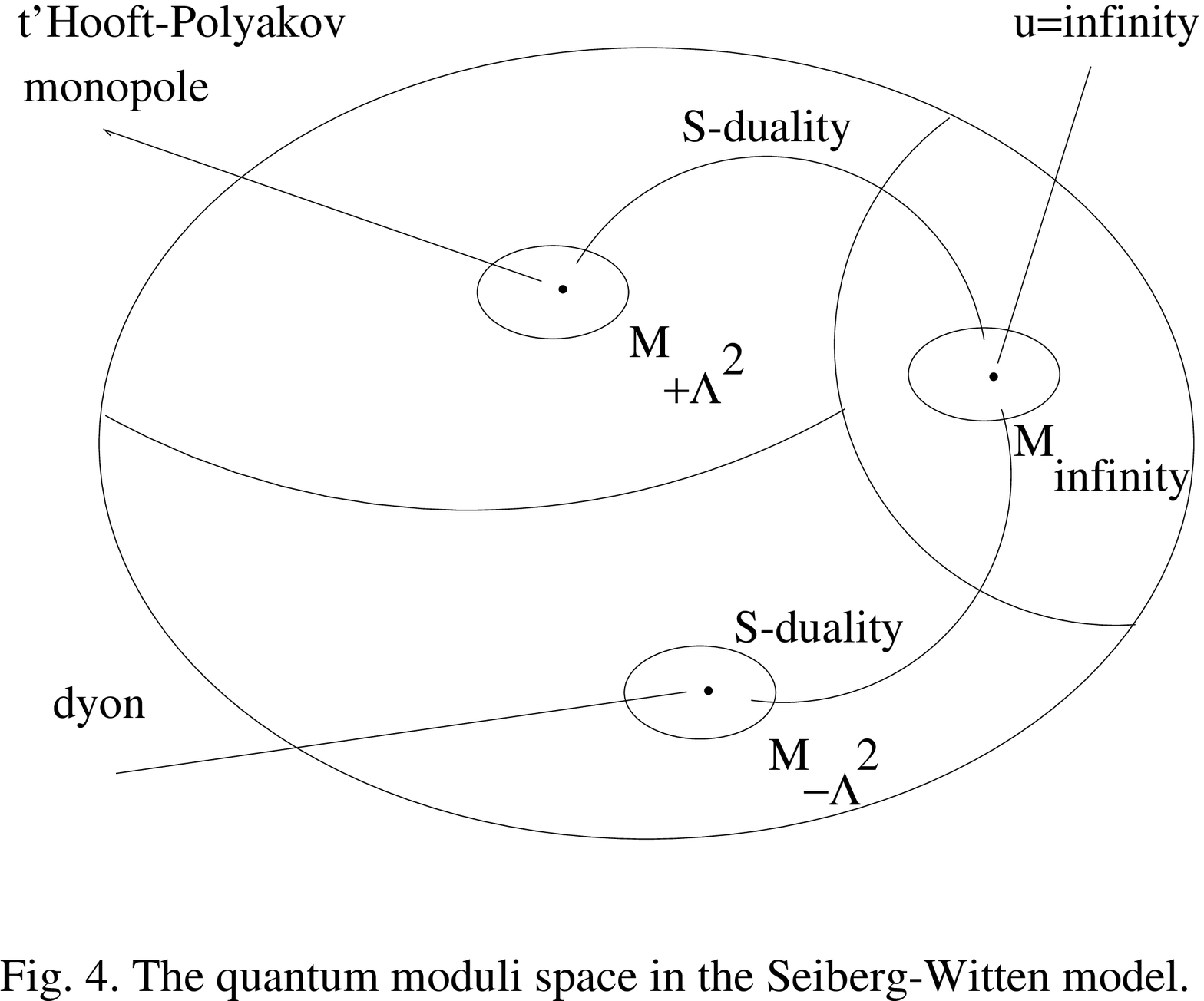}
}
\end{figure}

Consider, for simplicity, the famous Seiberg-Witten model~\cite{sw} whose 
fundamental action describes the purely gauge $N=2$ super-Yang-Mills theory
with the $SU(2)$ gauge group spontaneously broken to its $U(1)_e$ subgroup.
In the strong coupling region of the Coulomb branch, near a singularity in
the quantum moduli space where a BPS-like (t'Hooft-Polyakov) monopole becomes
massless, the Seiberg-Witten theory is just described by the S-dual $N=2$
supersymmetric QED. In particular, the t'Hooft-Polyakov monopole belongs to a
{\it magnetically} charged (HP) hypermultiplet $q^+_{HP}$ that represents the
non-perturbative degrees of freedom in the theory (Fig.~4). The HSS results of 
subsect.~3.2 imply that the HP hypermultiplet self-interaction in the vicinity
of the monopole singularity is regular in terms of the magnetically dual 
variables,
$$ \ck^{(+4)}\low{\rm Taub-NUT}({q^+}\low{HP}) 
=\fracmm{\l_{\rm dual}}{2}\left(
\sbar{q^+}\low{HP}{q^+}\low{HP}\right)^2~,\eqno(4.7)$$
i.e. it is given by the Taub-NUT (or KK-monopole).

On the other hand, from the type-IIA superstring (or M-theory) point of view,
the HP-hypermultiplet is just the zero mode of the open superstring stretching
between a magnetically charged D-6-brane and a D-4-brane. Therefore, it is the 
magnetically charged (HP) hypermultiplet that only survives in the effective
$4d$, $N=2$ gauge theory, after taking the local limit $\a'\to 0$ of the brane
configuration. According to the preceeding subsect.~4.1, the target space 
(NLSM) geometry governing the HP hypermultiplet self-interaction has to be the
Taub-NUT (or KK-monopole) again~!
\vglue.2in

\section{On the next-to-leading-order correction to the gauge LEEA}

The next-to-leading-order correction to the $N=2$ gauge LEEA in the Coulomb 
branch is governed by a real function of $W$ and $\bar{W}$ only, i.e. without
any dependence upon their $N=2$ superspace derivatives~\cite{wgr},
$$ \ch(W,\bar{W})=\ch_{\rm per.}(W,\bar{W})+\ch_{\rm non-per.}(W,\bar{W})~,
\eqno(5.1)$$
The exact function $\ch$ has to be S-duality invariant~\cite{henn}.

\begin{figure}
\vglue.1in
\makebox{
\epsfxsize=4in
\epsfbox{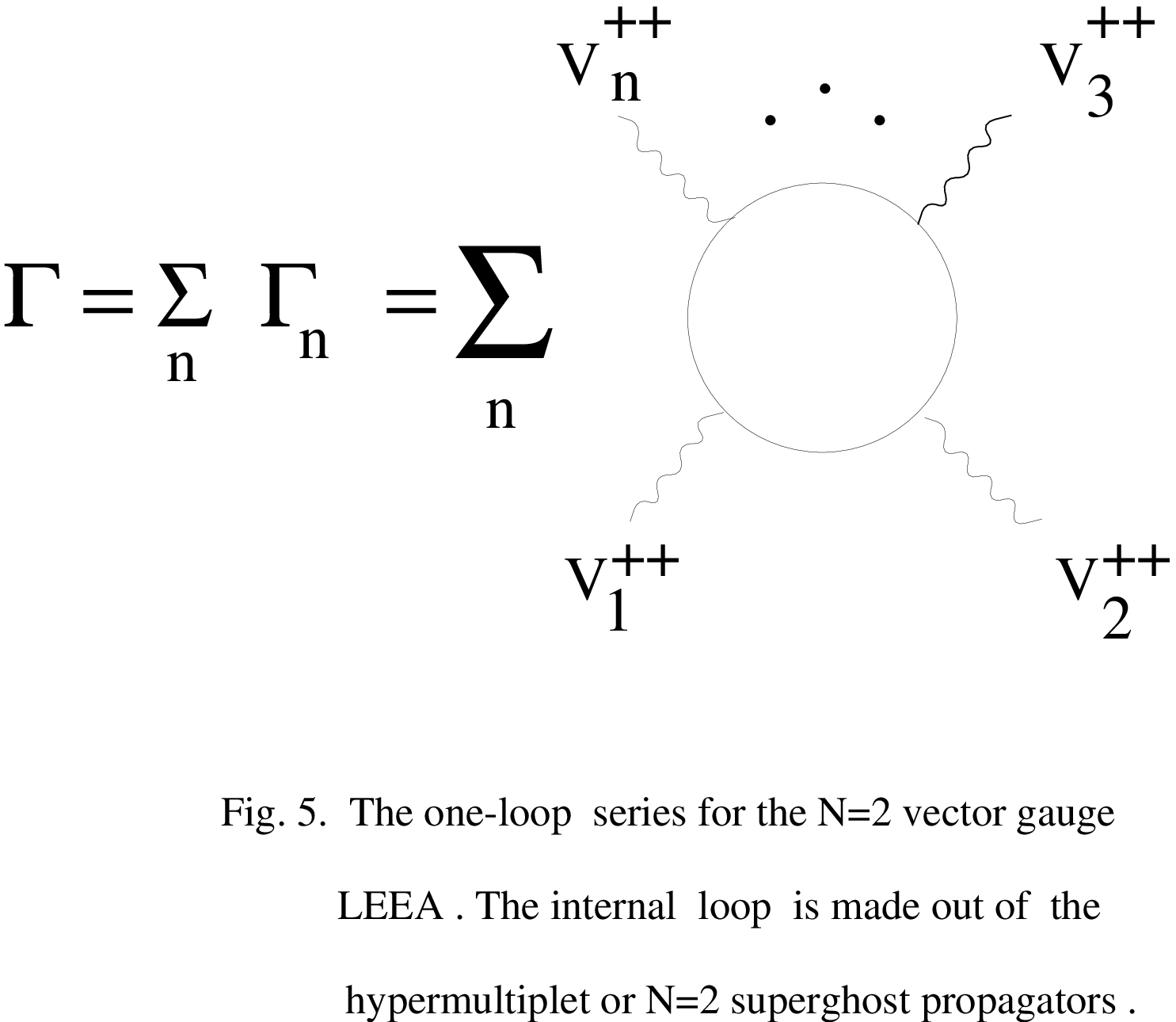}
}
\end{figure}

The one-loop contribution to $\ch_{\rm per.}$ is given by a sum of the $N=2$ 
HSS graphs schematically pictured in Fig.~5. The sum goes over the external
$V^{++}$-legs, whereas the loop consists of the $N=2$ matter (and $N=2$ ghost) 
superpropagators (sect.~3). $N=2$ ghosts contribute in very much the same
way as $N=2$ matter does, since the $N=2$ ghosts are also described in 
terms of the FS and HST hypermultiplets (with the opposite statistics of 
components) in the $N=2$ HSS~\cite{bb}. Because of the (abelian) gauge 
invariance, the result can only depend upon the abelian $N=2$ superfield 
strength $W$ and its conjugate $\bar{W}$ via eq.~(3.21). In fact, Fig.~5 also 
determines the one-loop perturbative contribution to the leading holomorphic 
LEEA, which appears as the anomaly associated with the non-vanishing 
central charges~\cite{bko}. The self-energy HSS supergraph with two external 
legs is the only one which is UV-divergent in Fig.~5. The IR-divergences 
of all the HSS graphs are supposed to be regulated by an IR-cutoff 
$\L$, which is proportional to the Seiberg-Witten scale introduced in sect.~2 
(the relative coefficient depends on the calculational scheme used, see e.g.,
ref.\cite{svkh}). One finds~\cite{bbi,svkh}
$$ \cf_{\rm 1-loop}\sim W^2\ln\fracmm{W^2}{\L^2}~,\qquad
\ch_{\rm 1-loop}\sim \ln^2\left(\fracmm{W\bar{W}}{\L^2}\right)~,\quad
\abs{W}\gg\L~,\eqno(5.2)$$
where we have identified the renormalization scale with $\L$. The numerical 
coefficients in front of the logarithms in eq.~(5.2) depend upon the content
 of the $N=2$ gauge theory under consideration, i.e. upon the data $(N_c,N_f)$.
The coefficient in front of the holomorphic contribution  is fixed by the
perturbative (one-loop) RG beta-function, i.e. it is proportional to
$(-2N_c+N_f)$ and is gauge-invariant. The coefficient in front of the logarithm
squared is proportional to $(-N_c+2N_f)$, in the $N=2$ super-Feynman gauge
which was actually used above~\cite{svkh}.

Eq.~(5.2) is the result of straightforward and manifestly $N=2$ supersymmetric
calculations in the $N=2$ HSS~\cite{bbi,bb,svkh}, and it agrees with the
standard arguments based on the perturbative $R$-symmetry and the integration 
of the associated chiral anomaly~\cite{sei,ds}. It is also straightforward to 
check (as I did) that there are no two-loop contributions to both 
$\cf_{\rm per.}$ and $\ch_{\rm per.}$, since all the relevant HSS graphs shown
in Fig.~6 do not actually contribute in the local limit. This conclusion is 
also in agreement with some recent calculations in terms of the $N=1$ 
superfields~\cite{ggruz}, as well as the general perturbative structure of the 
$N=2$ supersymmetric gauge field theories within the background-field method 
in $N=2$ HSS \cite{bko}. It does therefore seem to be conceivable that all 
the higher-loop contributions to $\ch_{\rm per.}(W,\bar{W})$ are absent too.

\begin{figure}
\vglue.1in
\makebox{
\epsfxsize=3in
\epsfbox{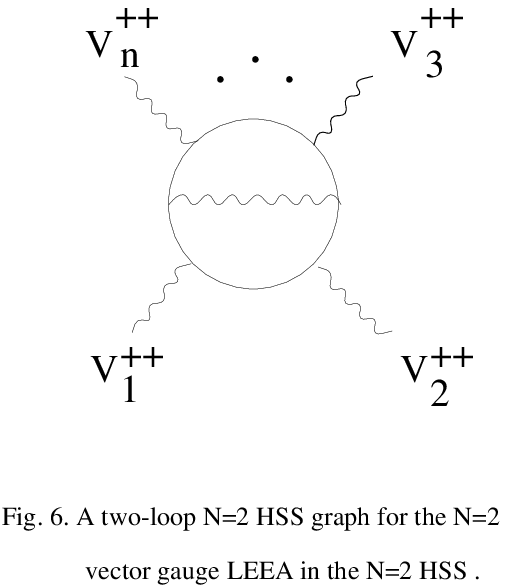}
}
\end{figure}

The exact result for the real function $\ch(W,\bar{W})$ is still unknown. There
is, however, an interesting proposal \cite{matone} that the exact function
$\ch(W,\bar{W})$ should satisfy a non-linear differential equation
$$ \pa\low{\bar{W}}\pa\low{W}\ln\left[\ch\pa\low{W}\pa\low{\bar{W}}\ln\ch
\right]=0~,\eqno(5.3)$$
which may be interpreted as a fully non-perturbative non-chiral superconformal
`Ward identity'.~\footnote{An explicit solution to eq.~(5.3) was also proposed
in ref.~\cite{matone}.} For instance, the leading one-instanton contribution 
in the pure $N=2$ gauge theory was already calculated in ref.~\cite{yung1}, 
and it does not vanish. The full non-perturbative contribution is not going to
be given by a sum over instanton contributions only, but it should also 
include (multi)anti-instanton and mixed (instanton-anti-instanton) 
contributions. The brane technology of sect.~4 might offer a direct procedure 
of calculating the exact next-to-leading-order contribution, by using the 
covariant action describing the M-theory 5-brane dynamics. The manifestly 
(world-volume) general coordinate invariant and supersymmetric action of the 
M-theory 5-brane is known~\cite{sch}, and it contains, in particular, a
Born-Infeld (BI) -type term and a Wess-Zumino (WZ) -type term. After being 
expanded in powers of derivatives, they yield the higher-derivative terms 
(in components). The latter are responsible for the exact form of the function
$\ch(W,\bar{W})$ in the effective LEEA of the M-theory 5-brane that should be 
related to the effective $N=2$ gauge field theory action.~\footnote{Similar 
remarks recently appeared in ref.~\cite{hlw} also.} However, as was argued in 
ref.~\cite{berk}, the actual results to be obtained from the brane technology 
may differ from that in the field theory, as regards the non-holomorphic terms
which are not fully protected by symmetries. The only alternative seems to be 
the use of the $N=2$ HSS in the instanton-type calculations, which is yet to 
be developed.

There are, however, some special cases when the non-perturbative corrections
to $\ch(W,\bar{W})$ vanish altogether. It just happens in the {\it scale
invariant} $N=2$ supersymmetric gauge field theories that cannot be (scale)
$\L$-dependent~\cite{ds}. This proposal is supported by the instanton
calculus~\cite{wales}. In the scale-invariant case, it is the one-loop
perturbative contribution to  $\ch(W,\bar{W})$ that is exact. It is easy to
check that the $\ch_{\rm per.}(W,\bar{W})$ of eq.~(5.2), in fact, does not
depend upon $\L$, since the real function $\ch(W,\bar{W})$ itself
is defined modulo the K\"ahler transformations
$$\ch(W,\bar{W})\to \ch(W,\bar{W})+f(W) +\bar{f}(\bar{W})~,
\eqno(5.4)$$
with an arbitrary holomorphic function $f(W)$ as a parameter.

In yet another scale-invariant $N=4$ supersymmetric Yang-Mills theory, that
amounts to the $N=2$ super-Yang-Mills multiplet minimally coupled to a 
hypermultiplet in the {\it adjoint} representation of the gauge group, both 
functions $\cf_{\rm int.}(W)$ and $\ch(W,\bar{W})$ vanish~\cite{svkh}.
\vglue.2in

\section{Hypermultiplet LEEA in the Higgs branch}

As was already mentioned in sect.~3, the most natural and manifestly $N=2$
supersymmetric description of hypermultiplets in the Higgs branch is provided 
by HSS in terms of the HST-type analytic superfield $\o$ of vanishing
$U(1)$ charge. The $N=2$ HSS is also the quite natural framework to address 
all possible symmetry breakings.

The free action of a single $\o$ superfield reads
$$ S[\o]=-\ha\int_{\rm analytic} \left(D_A^{++}\o\right)^2~.\eqno(6.1)$$
Similarly to the free action (3.11) for a $q^+$-type analytic superfield, the
action (6.1) also possesses the extended internal symmetry
$$SU(2)_A\otimes SU(2)_{\rm extra}~,\eqno(6.2)$$
where $SU(2)_A$ is the automorphism symmetry of $N=2$ supersymmetry algebra
(sometimes also called the $SU(2)_R$ symmetry). The extra $SU(2)$ symmetry of
eq.~(6.1) is a bit less obvious~\cite{ikz}~:
$$ \d\o=c^{--}D_A^{++}\o-c^{+-}\o~,\eqno(6.3)$$
where $c^{--}=c^{(ij)}u^-_iu^-_j$ and $c^{+-}=c^{(ij)}u^+_iu^-_j$, and
$c^{(ij)}$ are the infinitesimal parameters of $SU(2)_{\rm extra}\,$.

It is quite clear now that it is not possible to construct any non-trivial 
self-interaction in terms of the $U(1)$-chargeless superfield $\o$ alone,
simply because  any hyper-K\"ahler potential has $U(1)$ charge $(+4)$.
Hence, when $N=2$ supersymmetry and the $SU(2)_A$ internal symmetry are not
broken, one gets the well-known result~\cite{sw}:
$$\ck^{(+4)}\low{\rm Higgs}(\o)=0~,\eqno(6.3)$$
i.e. the induced hyper-K\"ahler metric in the fully $N=2$ supersymmetric Higgs
branch is flat.

It is, however, possible to break the internal symmetry (6.2) down to 
$$ U(1)_A\otimes SU(2)_{\rm extra}~,\eqno(6.4)$$
by introducing the so-called {\it Fayet-Iliopoulos} (FI) term
$$\VEV{D^{ij}}=\x^{ij}={\it const.}\neq 0~.\eqno(6.5)$$
This way of symmetry breaking still allows us to maintain control over the 
quantum hypermultipet LEEA because of the non-abelian internal symmetry (6.4). 
The only non-trivial hyper-K\"ahler potential, that is invariant w.r.t.
the symmetry (6.4) is given by~\cite{ikz}
$$\ck^{(+4)}\low{\rm EH}(\o)=-\,\fracmm{(\x^{++})^2}{\o^2}~,\eqno(6.6)$$
where $\x^{++}=\x^{ij}u^+_iu^+_j$. It is straightforward to deduce the
corresponding hyper-K\"ahler metric from eq.~(6.6) by using the procedure
already described in subsect.~3.2. One finds that the metric is
equivalent to the standard {\it Eguchi-Hanson} (EH) instanton
metric in four dimensions~\cite{giot}. The induced scalar potential was 
calculated in ref.~\cite{kup}.

It should be noticed that the hyper-K\"ahler potential (6.6) already implies
that $\VEV{\o}\neq 0$, so that we are in the Higgs branch indeed. Therefore, 
we now have to understand how a FI-term could be generated. Let's slightly 
generalize this problem by allowing non-vanishing vacuum 
expectation values for all the gauge-invariant bosonic components of the 
abelian $N=2$ superfield strength $W$,
$$ \VEV{W}=\left\{~ \VEV{A}=Z~,\quad \VEV{F_{\m\n}}=n_{\m\n}~,\quad
\VEV{\vec{D}}=\vec{\x} ~~\right\}~,\eqno(6.7)$$
where all the parameters $(Z,n_{\m\n},\vec{\x})$ are constants. Generally
speaking, it amounts to the {\it soft} $N=2$ supersymmetry breaking~\cite{amz}.
 We already know about the physical meaning of $Z$ --- it is just the central 
charge or 
the related gauge-invariant quantity $u\sim\VEV{\tr\,A^2}$, that parametrize 
the quantum moduli space of vacua. The central charge can be naturally 
generated via the standard Scherk-Schwarz mechanism of dimensional reduction 
from six dimensions~\cite{ikz}. Similarly, $n_{\m\n}\neq 0$ can be interpreted
as a {\it toron} background after replacing the effective spacetime $R^{1+3}$ 
by a hypertorus $T^{1+3}$ and imposing t'Hooft's twisted boundary 
conditions~\cite{toot}. The $\vec{\x}\neq 0$ is just a FI term.

The brane technology helps us to address the question of {\it dynamical} 
generation of both $n_{\m\n}$ and $\vec{\x}$ in a very geometrical way: namely,
one should deform the brane configuration of Fig.~3 by allowing the branes
to {\it intersect at angles} instead of being parallel~! Indeed, the vector
$\vec{w}=(x^7,x^8,x^9)$ is the same in Fig.~3 for both (NS) solitonic 5-branes.
Its non-vanishing value 
$$\vec{\x}=\vec{w}_1-\vec{w}_2\neq 0 \eqno(6.8)$$
effectively generates the FI term. Similarly, when allowing the D-4-branes to
intersect at angles, some non-trivial values of $\VEV{F_{\m\n}}=n_{\m\n}\neq 0$
are generated~\cite{dtoron}.

Since the LEEA of BPS branes is governed by a gauge theory, it does not seem to
 be very surprising that torons can also be understood as the BPS bound states
of certain D-branes in the field theory limit $\a'\to 0$
(or $M_{\rm Planck}\to\infty$)~\cite{dtoron}. Moreover, torons generate 
a gluino condensate~\cite{zhit}
$$ \VEV{\l^i\l^j}=\L^3(\x^2)^{ij}~,\quad \x^{ij}\sim \d^{ij}\exp\left(-\,
\fracmm{2\p^2}{g^2}\right)~,\eqno(6.9)$$
where $\vec{\x}\sim \{\x^{ij}\}$ have to be constant~\cite{itep}, and they can
be identified with the FI term by $N=2$ supersymmetry.

Finally, it is also quite useful to understand the origin of the 
hypermultiplet EH-type self-interaction in the Higgs branch from the viewpoint
of brane technology. It is worth mentioning here that the D-4-branes can also
end on the D-6-branes (in the type-IIA picture) so that these  D-4-branes
actually support hypermultiplets, not $N=2$ vector multiplets~\cite{wi}. It
results in another hyper-K\"ahler manifold $Q$ that has different topology 
$\sim S^3/Z_2$ in its spacial infinity. It is now enough to mention that the
EH-instanton is the only hyper-K\"ahler manifold having this topology among
the four-dimensional ALF spaces~!
\vglue.2in

\newpage

\section{Conclusion}

Though being very different, all the main three approaches considered above 
and depictured in Fig.~1, namely, 
\begin{itemize}
\item[(i)] instanton calculus, 
\item[(ii)] Seiberg-Witten approach and M theory (=brane technology),
\item[(iii)] harmonic superspace,
\end{itemize}
lead to the {\it consistent} results, as regards the leading terms in the LEEA
of the $4d$, $N=2$ 
supersymmetric gauge theories. The third (superspace) approach was mostly 
discussed in this paper, since it seems to be underrepresented in the current
literature. There is no unique universal method to handle all the problems 
associated with the $4d$ gauge theories in the most natural and easy way; in 
fact, each approach has its own advantages and disadvantages. For example, in 
the Seiberg-Witten approach, the physical information is encoded in terms of 
functions defined over the quantum moduli space whose modular group is 
identified with the duality group. The very existence of this approach is 
crucially dependent upon knowing exactly the perturbative limits of the gauge 
theory where, in its turn, the HSS approach is very efficient. At the same 
time, the HSS approach itself cannot be directly applied to address truly 
non-perturbative phenomena yet. It can, however, when being combined with the 
strong-weak couling duality (=S-duality). In its turn, the instanton calculus 
is very much dependent upon applicability of its own basic assumptions. It is 
not manifestly supersymmetric at any rate, if it is supersymmetric at all, and
it sometimes needs an additional input too. On the other hand, though being 
geometrically very transparent, the recently developed (M theory) brane 
technology has a rather limited analytic support by now, and its applications
are limited so far to those terms in the LEEA which are protected by $N=2$ 
supersymmetry, i.e. either holomorphic or analytic ones, Hence, a care should 
be excercised in order to play safely with it. I believe, it is a combination 
of all the methods available that has the strongest potential for a further 
progress, and that simultaneously teaches us how to proceed with each 
particular approach.

I would like to conclude with a few comments about $N=2$ supersymmetry breaking
and confinement, in order to indicate on a possible importance of the exact
hypermultiplet low-energy effective action towards a solution to these 
problems. Indeed, it seems to be quite natural to take advantage of the 
existence of exact solutions to the low-energy effective action in $N=2$ 
supersymmetric gauge field theories, and apply them to the old problem of 
color confinement in QCD. In fact, it was one of the main motivations in the 
original work of Seiberg and Witten~\cite{sw}. The most attractive mechanism 
for color confinement is known to be the dual Meissner effect or the dual 
(Type II) superconductivity \cite{mant}. It takes three major steps to connect
an ordinary BCS superconductor to the simplest Seiberg-Witten model in quantum
field theory: first, define a {\it relativistic} version of the 
superconductor, known as the (abelian) Higgs model in field theory, second, 
introduce a {\it non-abelian} version of the Higgs model, known as the 
Georgi-Glashow model, and, third, $N=2$ {\it supersymmetrize} the 
Georgi-Glashow model in order to get the Seiberg-Witten model~\cite{sw}. Since
the t'Hooft-Polyakov monopole of the Georgi-Glashow model belongs to a (HP)
hypermultiplet in its $N=2$ supersymmetric (Seiberg-Witten) generalisation,
it is quite natural to explain confinement as the result of a monopole
condensation (= the dual Meissner effect as a consequence of the dual 
Higgs effect), i.e. a non-vanishing vacuum expectation value for the 
magnetically charged (dual Higgs) scalars belonging to the HP hypermultiplet. 
Of course, it is only possible after (or simultaneously with) $N=2$ 
supersymmetry breaking.

Exact solutions to the low-energy effective action in quantum gauge field 
theories are only available in $N=2$ supersymmetry, and neither in $N=1$ 
supersymmetry nor in the bosonic QCD. Hence, on the one side, it is the $N=2$ 
supersymmetry that crucially simplifies an evaluation of the low-energy 
effective action. However, on the other side, it is the same $N=2$ 
supersymmetry that is so obviously incompatible with phenomenology e.g., 
because of equal masses of bosons and fermions inside $N=2$ supermultiplets 
(it also applies to any $N\geq 1$ supersymmetry), and the non-chiral nature 
of $N=2$ supersymmetry (e.g. `quarks' appear in real representations of the 
gauge group). Therefore, if we believe in the $N=2$ supersymmetry, we should 
find a way of judicious $N=2$ supersymmetry breaking. The same dual Higgs 
mechanism may also be responsible for the chiral symmetry breaking and the 
appearance of the pion effective Lagrangian if the dual Higgs field has flavor
charges also~\cite{sw}. In fact, Seiberg and Witten used a mass term for the 
$N=1$ chiral multiplet, which is a part of the $N=2$ vector multiplet, in 
order to {\it softly} break $N=2$ supersymmetry to $N=1$ supersymmetry. As a 
result, they found a non-trivial vacuum solution with a monopole condensation 
and, hence, a confinement. The weak point of their approach is an {\it ad hoc}
assumption about the existence of the mass gap, i.e. the mass term itself. It
would be nice to derive the mass gap from the fundamental theory instead of
postulating it. 

The $N=2$ supersymmetry can be broken either softly or spontaneously, if one
wants to preserve the benefits of its presence (e.g. for the full control
over the low-energy effective action) at high energies. The general analysis 
of the {\it soft} $N=2$ supersymmetry breakings in the $N=2$ supersymmetric 
QCD was given by Alvarez-Gaum\'e, Mari\~no and Zamora~\cite{amz}.~\footnote{
See e.g., ref.~\cite{dms} for a similar analysis in $N=1$ supersymmetric gauge
field theories.} The soft $N=2$ supersymmetry  breaking is most naturally done
by using the FI-terms, as in ref.~\cite{amz}. Though being pragmatic, the soft
$N=2$ supersymmetry breaking has a limited predictive power because of many 
parameters, whose number, however, is significatly less than that in the
$N=1$ case. Hence, it should make sense to search for the patterns of 
{\it spontaneous} $N=2$ supersymmetry breaking, where the non-vanishing 
FI-terms would appear as stationary solutions to the dynamically generated 
scalar potential. This would mean the existence of a non-supersymmetric
vacuum solution for the $N=2$ supersymmetric scalar potential at the level of
the low-energy effective action in $N=2$ gauge theories. Since the $N=2$
supersymmetry remains unbroken for any exact Seiberg-Witten solution in the
gauge sector, we should consider the induced scalar potentials in the 
hypermultiplet sector of an $N=2$ gauge theory. Indeed, given the non-trivial 
kinetic terms in the hypermultiplet low-energy effective action to be 
represented by the (hyper-K\"ahler) non-linear sigma-model, in a presence of 
non-vanishing central charges they also imply  a non-trivial hypermultiplet 
scalar potential whose form is entirely determined by the hyper-K\"ahler 
metric of the kinetic terms and $N=2$ supersymmetry. Though it is not easy
to search for the most general solutions with spontaneously broken $N=2$
supersymetry because of complications associated with HSS and hyper-K"ahler
geometry, our examples demonstrate the richness of possible solutions.

\section*{Acknowledgements}

I would like to thank the Physics Department of the University of Maryland at 
College Park, the Physics Department of Pennsylvania State University, and the
Lyman Laboratory of Physics in Harvard University for a kind hospitality 
extended to me during the completion of this work. I am also grateful to Luis 
Alvarez-Gaum\'e, Jonathan Bagger, Diego Bellisai, Joseph Buchbinder, Norbert
Dragon, Jim Gates, Marc Grisaru, Andrei Johansen, Wolfgang Lerche, 
Marcus Luty, Andrei Mironov, Alexei Morozov, Burt Ovrut, Norisuke Sakai, 
John Schwarz, Dima Sorokin, Cumrun Vafa and Alexei Yung for useful discussions.

\end{document}